\documentclass[aps,prd,twocolumn,superscriptaddress,nofootinbib,floatfix]{revtex4-2}
\usepackage{amsmath,amssymb,graphicx,bm,mathrsfs}
\usepackage[colorlinks=true,linkcolor=blue,citecolor=blue,urlcolor=blue]{hyperref}
\usepackage{orcidlink}
\usepackage{placeins}

\newcommand{\scri}{\mathscr{I}^+}
\newcommand{\Sw}{{}_{-2}S}
\newcommand{\Yw}{{}_{-2}Y}
\newcommand{\dd}{\mathrm{d}}

\begin{document}

\title{Kerr greybody factors in spherical multipoles:\\
time-domain scattering and mode mixing in numerical-relativity ringdowns}

\author{Takol Tangphati \orcidlink{0000-0002-6818-8404}}
\email[]{takoltang@gmail.com}
\affiliation{School of Science, Walailak University, Thasala, \\Nakhon Si Thammarat, 80160, Thailand}
\affiliation{Research Center for Theoretical Simulation and Applied Research in Bioscience and Sensing, Walailak University, Thasala, Nakhon Si Thammarat 80160, Thailand}

\author{Ayan Banerjee \orcidlink{0000-0003-3422-8233}}
\email[]{ayanbanerjeemath@gmail.com}
\affiliation{Astrophysics Research Centre, School of Mathematics, Statistics and Computer Science, University of KwaZulu-Natal, Private Bag X54001, Durban 4000, South Africa}
\affiliation{Astrophysics and Cosmology Research Unit, School of Mathematics, Statistics and Computer Science, University of KwaZulu-Natal, Private Bag X54001, Durban 4000, South Africa}

\date{\today}

\begin{abstract}
Kerr perturbations separate in spin-weighted spheroidal harmonics, whereas numerical-relativity waveforms are
decomposed in spherical harmonics. Greybody-factor models of the ringdown have so far neglected the resulting mode
mixing. We quantify it for $m=2$. Time-domain evolutions of the Teukolsky equation on hyperboloidal slices reproduce the
frequency-domain greybody factors of the dominant mode to within $1.3\times10^{-4}$ for black-hole spins up to $0.99$,
and the relative phases of the horizon-normalised reflected amplitudes of the tested modes to about $10^{-2}$\,rad. For
equal source amplitudes, the $L=3$ wave leaking into the spherical $\ell=2$ channel overtakes the direct $L=2$
reflection at $M\omega=0.63$--$0.91$ for spins $0.3$--$0.99$, and exceeds it by factors of $1.8$--$84$ at $M\omega=1$.
For the source ratios inferred from nine numerical-relativity mergers, the leak reaches $8$--$38\%$ of the direct
reflection at the top of the usable band in eight of them ($6$--$37\%$ in the six runs with an extrapolation-order comparison, on
the bands it supports). In the $(3,2)$ multipole the median ratio of leaked to direct amplitude
ranges from $0.04$ to $7.9$. Quasinormal-mode fits recover the predicted $(2,2,0)$ admixture within the fit spread in
eight of the nine mergers. Frequency-domain demixing with a fitted mixing strength disfavours the single-mode
description in seven of them, beyond the 95th percentile of a null distribution; with the strength fixed at the
prediction it does so in four. In Schwarzschild the sign of the imaginary part of the
Teukolsky--Starobinsky constant selects the parity of the reflected wave; we adopt its continuation to Kerr as a
prescription. Greybody-factor models of the $(3,2)$ multipole must include the mixing.
\end{abstract}

\maketitle

\section{Introduction}
\label{sec:intro}

The ringdown of a binary black-hole merger is commonly described as a superposition of quasinormal modes (QNMs) of the
remnant Kerr black hole~\cite{Vishveshwara:1970zz,Detweiler:1980gk,Leaver:1985ax,Kokkotas:1999bd,Nollert:1999ji,Berti:2009kk}.
Measuring several QNMs tests the Kerr nature of the remnant (black-hole
spectroscopy)~\cite{Dreyer:2003bv,Berti:2005ys,LIGOScientific:2016lio,Brito:2018rfr,Isi:2019aib}, but QNM fits close to the
merger depend on the treatment of overtones and nonlinearities~\cite{Giesler:2019uxc,Mitman:2022qdl,Cheung:2022rbm},
and the QNM spectrum itself is unstable under small perturbations of the
potential~\cite{Nollert:1996rf,Jaramillo:2020tuu,Cheung:2021bol}. Recently an alternative description has been developed in which the spectral amplitude of the
post-merger signal is governed by the greybody factor~\cite{Hawking:1975vcx,Page:1976df,Starobinskil:1974nkd,Teukolsky:1974yv} of the remnant, i.e.\ by the energy transmissivity (or,
equivalently, reflectivity) of the effective potential barrier, multiplied by a smooth source
term~\cite{Oshita:2022pkc,Oshita:2023cjz,Okabayashi:2024qbz,Rosato:2025ulx,Rosato:2026apq}. In the extreme mass-ratio
limit the spectral amplitude of the ringdown follows the reflectivity closely~\cite{Oshita:2023cjz}; for
comparable-mass mergers simulated in numerical relativity (NR), greybody-factor models fit the frequency-domain
ringdown amplitude of the dominant multipole~\cite{Okabayashi:2024qbz,Rosato:2025ulx}; and they have been turned into
new tests of general relativity~\cite{Rosato:2026apq}. A key motivation is stability: unlike the QNM spectrum,
greybody factors change only slightly under small perturbations of the potential~\cite{Rosato:2024arw,Oshita:2024fzf}.
Greybody factors and QNMs are nevertheless closely related~\cite{Konoplya:2024lir,Huang:2025rxx}.
The subject has recently been reviewed in Ref.~\cite{Konoplya:2026qql}.

All of these models describe a spherical-harmonic multipole $h_{\ell m}$ of the strain, which is what NR codes and
gravitational-wave analyses use, with the reflection amplitude of the Kerr spheroidal mode with the same $(\ell,m)$.
For a rotating remnant, however, perturbations separate in spin-weighted \emph{spheroidal} harmonics
${}_{-2}S_{Lm}(\theta;a\omega)$, which depend on frequency, and each spherical multipole receives contributions from
several spheroidal modes~\cite{Berti:2014fga,Cook:2014cta}. In the QNM description this spherical--spheroidal mixing
is well established: the $(3,2)$ multipole of NR ringdowns contains a prominent component oscillating at the $(2,2,0)$
QNM frequency~\cite{Buonanno:2006ui,Kelly:2012nd,London:2014cma,Berti:2014fga}. In the greybody-factor description it has been neglected so far. In
Ref.~\cite{Rosato:2025ulx} the mixing is expected to be small in the frequency range of interest, while
Ref.~\cite{Rosato:2026apq} names its inclusion as an important direction, and the review~\cite{Konoplya:2026qql}
lists mode mixing among the open limitations of greybody-based ringdown templates. In waveform modelling, Mehta
\emph{et al.}~\cite{KumarMehta:2019izs} removed and restored the mixing in the $(3,2)$ and $(4,3)$ multipoles of
nonspinning hybrid waveforms, using mixing coefficients evaluated at the fundamental QNM frequency. A greybody-factor
model needs instead the coefficients at real frequency, $\mu(a\omega)$, combined with the Kerr reflection amplitudes.

In this paper we quantify spherical--spheroidal mixing in greybody-factor models of Kerr ringdowns for $m=2$ and
test it against NR. The reflection amplitudes are computed in the frequency domain (FD)~\cite{BHPToolkit,Sasaki:2003xr}.
We validate them with an independent time-domain (TD) scattering experiment: we evolve the Teukolsky
equation~\cite{Teukolsky:1973ha} in 2+1 dimensions on the hyperboloidal, horizon-penetrating foliation of Kerr
introduced in Refs.~\cite{Racz:2011qu,Harms:2013ib}, send in an ingoing gravitational-wave packet with a prescribed
spherical multipole, and read the full angular profile of the field at the event horizon and at future null infinity
$\scri$, without any outer boundary. Time-domain evolutions of spin-$s$ fields on Kerr on this foliation have been used
to study late-time tails~\cite{Harms:2013ib,Csukas:2019kcb} and superradiant energy gains of frequency-tuned
packets~\cite{Csukas:2021sia}; here the scattered wave is resolved in frequency and in spheroidal mode, which extends the
1+1 reflectivity measurements of Ref.~\cite{Oshita:2024fzf} to rotating black holes, where mixing cannot be captured in
1+1 dimensions. The evolution serves as a validation of the FD amplitudes and as the platform for our work beyond
Kerr.

Our results are as follows.
(i) For the dominant mode the TD greybody factors agree with the FD values to $\le1.3\times10^{-4}$ for
$|M\omega|\le0.95$ and black-hole spins up to $\chi=0.99$ (where only $L=2$ is tested), including the superradiant
band; subdominant modes agree to $\lesssim10^{-2}$ where $|\Gamma_L|>10^{-3}$, and the relative phases of the
horizon-normalised reflected amplitudes of different modes agree to $\lesssim10^{-2}$\,rad. The FD amplitude of the
dominant mode also agrees with the public data of Ref.~\cite{Rosato:2026apq} to $\le4.5\times10^{-4}$ at the four spins compared
(Sec.~\ref{sec:validation}). The FD amplitudes validated in this way are used in the rest of the paper; at
$M\omega=1$--$1.2$ they rest on the FD calculation alone.
(ii) The phase of the energy-normalised reflection amplitude depends on the sign of the imaginary part of the
Teukolsky--Starobinsky constant. At $\chi=0$ this sign selects the parity of the reflected wave, as is implicit in the
classical relation between the Teukolsky and Regge--Wheeler/Zerilli
equations~\cite{Chandrasekhar:1975nkd,Chandrasekhar:1985kt,Futterman:1988ni,Dolan:2008kf}; we verify this numerically.
For non-precessing binaries the $\ell+m$ odd multipoles correspond to the opposite sign to the $\ell+m$ even ones; at
$\chi\neq0$ we adopt the continuation of this assignment as a prescription. The sign is observable only when the
relative phases of the sources are fixed (Sec.~\ref{sec:parity}).
(iii) For equal source amplitudes, the reflected $L=3$ wave leaking into the spherical $\ell=2$ channel overtakes the
direct $L=2$ reflection at $M\omega=0.63$, $0.66$, $0.76$ and $0.91$ for $\chi=0.3$, $0.7$, $0.9$ and $0.99$, and
exceeds it by factors of $1.8$--$84$ at $M\omega=1$. For the sources inferred from NR the leak reaches $8$--$38\%$ of
the direct reflection at the top of the usable band of $\tilde h_{22}$ in eight runs and $66\%$ in the ninth, whose band
is not supported at those frequencies by a resolution comparison. In the $(3,2)$ multipole the leaked $(2,2)$
reflection dominates below ${\rm Re}\,\omega_{220}$ in six of nine runs, because the $(2,2)$ source is much stronger
(Secs.~\ref{sec:mixing} and~\ref{sec:nr}).
(iv) In nine SXS~\cite{Boyle:2019kee,Scheel:2025jct} mergers with mass ratios $1$--$8$, QNM fits reproduce the predicted
mixing coefficient. FD demixing with a fitted mixing strength disfavours the single-mode description beyond the 95th
percentile of a null distribution built from all $72$ pairings of unrelated multipoles in seven runs; with the strength
fixed at the predicted value it does so in four, marginally in a fifth (94th percentile), and not in SXS:BBH:1124.
Mock data with the measured waveform differences show that the analysis recovers the predicted strength, but they are
much cleaner than the NR spectra and do not calibrate these numbers (Sec.~\ref{sec:nr}). The data also delimit what the spectra can say: the $(3,2)$ reflection amplitude is
preferred over alternatives in the five runs with $\chi_f\le0.83$ but not at higher spin, and the $(2,2)$ spectrum
above $M_f\omega\simeq0.85$ carries features in several unequal-mass runs that mixing does not explain
(Sec.~\ref{sec:excess}). For the example fit band of Ref.~\cite{Rosato:2025ulx}, $M\omega=0.35$--$0.75$, the leak in
$\tilde h_{22}$ of the three near-equal-mass mergers exceeds the smooth-source fit residuals only above
$M_f\omega\simeq0.72$, $0.81$ and $0.96$.

We use geometric units $G=c=1$, the time dependence $e^{-i\omega t+im\phi}$, and set the black-hole mass $M=1$ unless
stated otherwise; radii are then measured in units of $M$. The dimensionless spin is $\chi=a/M$. Since only $m=2$ is
studied, quantities of spheroidal mode $L$ carry the single index $L$ ($R_L$, $\sigma_L$, $\Gamma_L$); spherical
multipoles carry $(\ell,m)$, as in $h_{\ell m}$. SXS denotes the Simulating eXtreme Spacetimes collaboration and its
catalog, Lev its resolution levels, BMS the Bondi--Metzner--Sachs frame freedom of asymptotic waveforms, and CCE
Cauchy-characteristic extraction.

\section{Greybody factors and reflection amplitudes of Kerr}
\label{sec:setup}

\subsection{Teukolsky equation and asymptotic amplitudes}
Outgoing radiation is described by the Teukolsky field of spin weight $s=-2$, $\Psi=\rho^{-4}\psi_4$ with
$\rho=-1/(r-ia\cos\theta)$~\cite{Teukolsky:1973ha,Press:1973zz,Chandrasekhar:1985kt}. For a mode $e^{-i\omega t+im\phi}\Phi(r)\,\Sw_{Lm}(\theta;a\omega)$
the radial function obeys
\begin{multline}
\Delta^{2}\frac{\dd}{\dd r}\!\left(\Delta^{-1}\frac{\dd \Phi}{\dd r}\right)\\
+\left(\frac{K^2+4i(r-M)K}{\Delta}-8i\omega r-\lambda\right)\Phi=0,
\label{eq:radial}
\end{multline}
with $\Delta=r^2-2Mr+a^2$, $K=(r^2+a^2)\omega-am$, and separation constant $\lambda=A+a^2\omega^2-2am\omega$, where
$A$ is the angular eigenvalue. The solution that is purely ingoing at the horizon behaves as
\begin{equation}
\Phi^{\rm in}\to
\begin{cases}
\Delta^{2}e^{-ikr_*}, & r\to r_+,\\
B^{\rm inc}r^{-1}e^{-i\omega r_*}+B^{\rm ref}r^{3}e^{i\omega r_*}, & r\to\infty,
\end{cases}
\end{equation}
with $k=\omega-m\Omega_H$, $\Omega_H=a/(2Mr_+)$, and the tortoise coordinate $r_*$.

The incident and reflected energy fluxes involve the Teukolsky--Starobinsky
constant~\cite{Teukolsky:1974yv,Chandrasekhar:1985kt,Hughes:1999bq,Sasaki:2003xr},
\begin{multline}
|C|^2=\left[(\lambda+2)^2+4a\omega m-4a^2\omega^2\right]\\
\times\left(\lambda^2+36a\omega m-36a^2\omega^2\right)\\
+(2\lambda+3)\left(96a^2\omega^2-48a\omega m\right)+144\omega^2(M^2-a^2).
\label{eq:C2}
\end{multline}
Following Refs.~\cite{Rosato:2025ulx,Rosato:2026apq}, we define the energy-normalised reflection amplitude and the
greybody factor of mode $L$ by
\begin{equation}
R_L=\frac{C_L\,B^{\rm ref}_L}{(2M\omega)^4B^{\rm inc}_L},\qquad
|R_L|^2=1-\frac{\omega-m\Omega_H}{\omega}\,\Gamma_L .
\label{eq:RG}
\end{equation}
In the superradiant regime $0<\omega<m\Omega_H$ one has
$|R_L|>1$~\cite{Starobinskil:1974nkd,Teukolsky:1974yv,Brito:2015oca}. The modulus of $C_L$ follows from
Eq.~\eqref{eq:C2}; its phase is discussed in Sec.~\ref{sec:parity}.

\subsection{Spherical basis}
\begin{figure*}[t]
\includegraphics[width=\textwidth]{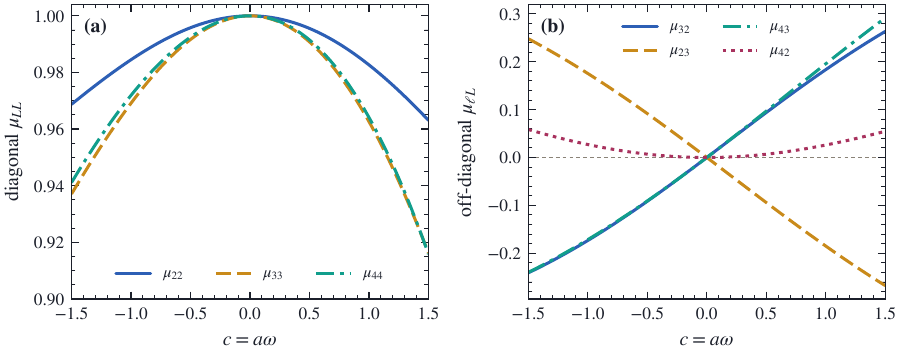}
\caption{\label{fig:mixing}Spherical--spheroidal mixing coefficients $\mu_{\ell L}=\langle\Yw_{\ell 2}|\Sw_{L2}(c)\rangle$
for $m=2$ as functions of the spheroidicity $c=a\omega$, with the phase of $\Sw_{L2}$ fixed by $\mu_{LL}>0$ and
$\Yw_{\ell2}$ in the convention of Ref.~\cite{Berti:2014fga}. (a) Diagonal coefficients. (b) Off-diagonal coefficients
coupling neighbouring multipoles ($\mu_{32}$, $\mu_{23}$, $\mu_{43}$) and next-to-neighbouring ones ($\mu_{42}$).}
\end{figure*}

The spheroidal harmonics expand in spherical harmonics of the same spin weight and azimuthal number,
\begin{equation}
\Sw_{Lm}(\theta;c)=\sum_{\ell}\mu_{\ell L}(c)\,\Yw_{\ell m}(\theta),
\label{eq:mu}
\end{equation}
where $\mu_{\ell L}=\langle \Yw_{\ell m}|\Sw_{Lm}\rangle$,
and $c=a\omega$. Both families are normalised to unit norm on the sphere; $\Yw_{\ell m}$ follows the convention of
Ref.~\cite{Berti:2014fga} and the phase of $\Sw_{Lm}$ is fixed by requiring $\mu_{LL}>0$, which fixes the signs of
the off-diagonal $\mu_{\ell L}$ used below. For real $c$ the coefficients are real, $\mu_{LL}=1-\mathcal{O}(c^2)$, and
$\mu_{L\pm1,L}=\mathcal{O}(c)$~\cite{Berti:2014fga,Cook:2014cta}. If a perturbation source excites spheroidal mode
$L$ with frequency-domain amplitude $\sigma_L(\omega)$ and the outgoing wave is reflected by the barrier, the spherical
multipole of the strain at infinity is
\begin{equation}
\tilde h_{\ell m}(\omega)=\sum_L \mu_{\ell L}(a\omega)\,R_L(\omega)\,\sigma_L(\omega),
\label{eq:response}
\end{equation}
where the frequency-dependent factor relating $\psi_4$ to the strain is absorbed into $\sigma_L$.
Figure~\ref{fig:mixing} shows the coefficients relevant here. The off-diagonal ones grow almost linearly with $c$ and
reach $|\mu_{32}|\simeq0.25$ at $|c|=1.5$; $\mu_{23}$ and $\mu_{32}$ have opposite signs, and $\mu_{42}$ is of second order.

Single-mode greybody models keep only the $L=\ell$ term. We call
$\mathcal{K}_{\ell\leftarrow L}=|\mu_{\ell L}R_L|/|\mu_{\ell\ell}R_\ell|$ the leakage kernel: it is the ratio of the
leaked to the direct contribution to channel $\ell$ for equal source amplitudes.

\section{Time-domain scattering on Kerr}
\label{sec:method}

\subsection{Hyperboloidal evolution}
Time-domain solutions of the Teukolsky equation in 2+1 dimensions have a long
history~\cite{Krivan:1997hc,Pazos-Avalos:2004uyd,Sundararajan:2007jg}. Hyperboloidal slices that reach $\scri$ avoid
artificial outer boundaries~\cite{Zenginoglu:2007jw,Zenginoglu:2011jz}. We evolve the $s=-2$ Teukolsky equation for
fixed $m$ in 2+1 dimensions on the R\'acz--T\'oth foliation~\cite{Racz:2011qu,Harms:2013ib}. Following
Ref.~\cite{Harms:2013ib}, Eqs.~(3)--(6), we pass from Boyer--Lindquist coordinates to ingoing Kerr coordinates,
\begin{equation}
t_{\rm in}=t-r+\int\frac{r^2+a^2}{\Delta}\,\dd r,\qquad \tilde\phi=\phi+a\int\frac{\dd r}{\Delta},
\label{eq:ingoing}
\end{equation}
and then to the compactified radius $X$ and the hyperboloidal time $\tau$,
\begin{equation}
r=\frac{2X}{1-X^2},\qquad \tau=t_{\rm in}-\sqrt{1+r^2}+4M\ln\!\left(1-X^2\right).
\label{eq:foliation}
\end{equation}
Slices of constant $\tau$ extend from just inside the future horizon to future null infinity $\scri$ at $X=1$; both
boundaries are outflow boundaries, so no boundary conditions are needed. We evolve $\psi$, defined by
$\Psi_{\rm BL}=\Delta^{-s}r^{-1}\psi$, which is regular at both boundaries, with fourth-order finite differences and
fourth-order Runge--Kutta time integration. The coefficients of the equation were derived independently from the
Boyer--Lindquist Teukolsky operator; the characteristic speeds at the boundaries, $dX/d\tau=-0.04607$ at the horizon
and $0.0800$ at $\scri$ for $\chi=0$, agree with Ref.~\cite{Harms:2013ib}. The code is written in Python with a JAX
core~\cite{jax2018github} in double precision; a time-domain code on the same foliation is
Teukode~\cite{Harms:2014dqa}. The standard resolution is $801$ radial and $32$ polar points. The numerical scheme and
its convergence are described in Appendix~\ref{app:numerics}.

For the greybody factors we repeated the $\chi=0.9$, $\ell'=2$ scattering run with $1201$ radial points and,
separately, with $48$ polar points: over $0.3\le|M\omega|\le0.95$ the relative change of $\Gamma_2$ has a median of
$3\times10^{-7}$ in both cases. Its maximum is $3\times10^{-4}$ for the radial refinement, at $M\omega=0.3$, where
$\Gamma_2$ is smallest ($3\times10^{-3}$), and $6\times10^{-5}$ for the polar refinement, at $M\omega=0.62$. The $\ell=m=2$ fundamental QNM, extracted by the matrix-pencil method from the ringdown at a single polar
angle, agrees with Ref.~\cite{Stein:2019mop} to $4$--$5\times10^{-6}$ ($\chi=0$), $3$--$6\times10^{-4}$ ($\chi=0.5$) and
$1$--$8\times10^{-5}$ ($\chi=0.9$), the range covering the horizon and $\scri$. At $\chi=0.5$ the single-angle
extraction is limited by neighbouring modes; the scattering tests of Sec.~\ref{sec:validation} are more stringent. For the
configuration of Ref.~\cite{Harms:2013ib} with initial data that do not vanish at $\scri$ ($\chi=0$, $m=2$,
$\psi=0$ and $\partial_\tau\psi\propto e^{-w(X-X_0)^2/2}\,\Yw_{22}$ with the dimensionless parameters $w=3000$ and $X_0=0.8$ in the
compactified radius), the local power index over $\tau=400$--$900M$ gives the late-time
decay rates~\cite{Price:1971fb} $t^{-6.1}$ at the horizon and $u^{-5.05}$ at $\scri$, as in that reference, against
the asymptotic values $6$ and $5$.

\subsection{Incident wave packets}
The incident packet is an exact solution of the flat-space $s=-2$ Teukolsky equation that is purely ingoing and has a
single spherical multipole $\ell'$. With $v$ the ingoing null coordinate and $F$ an arbitrary profile,
\begin{align}
\ell'=2:&\quad \Psi=\frac{F(v)}{r},\nonumber\\
\ell'=3:&\quad \Psi=\frac{F'(v)}{r}-\frac{3F(v)}{r^2},\label{eq:indata}\\
\ell'=4:&\quad \Psi=\frac{F''(v)}{r}-\frac{7F'(v)}{r^2}+\frac{14F(v)}{r^3},\nonumber
\end{align}
each multiplied by $\Yw_{\ell'm}(\theta)$ (Appendix~\ref{app:data}). We use a Gaussian $F$ of width $2M$ centred at
$r_0=35M$. On a Kerr background these data contain a small outgoing component at order $M/r_0$, which reaches $\scri$
first and merges into the leading edge of the scattered signal [Fig.~\ref{fig:td}(b)]; the transforms include it. Its
effect is small: moving the launch radius to $r_0=25M$ changes $\Gamma_2$ at $\chi=0.7$ by at most $4\times10^{-5}$ and
the horizon-normalised reflected amplitude $Z^\infty_2/Z^H_2$ (Sec.~\ref{sec:readout}) by at most $2.6\times10^{-3}$ over
$0.3\le|M\omega|\le0.95$, and leaves the agreement with the FD greybody factor unchanged ($4\times10^{-5}$). Gating the
residue out of the record is not a clean alternative, because the low-frequency part of the reflected signal overlaps
it.

\subsection{Read-out and spectral analysis}
\label{sec:readout}
\begin{figure}[t]
\includegraphics[width=\columnwidth]{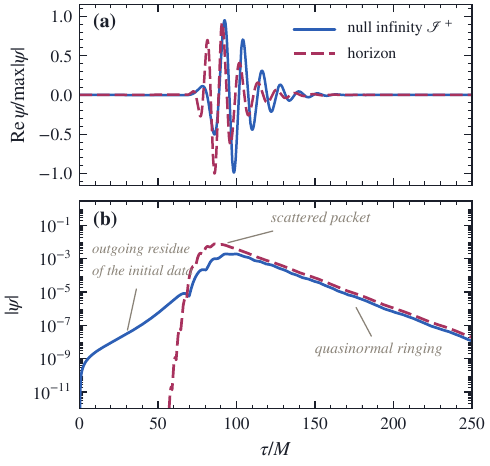}
\caption{\label{fig:td}Time-domain scattering of an $\ell'=2$, $m=2$ packet off a Kerr black hole with $\chi=0.7$, read
at $\theta=\pi/2-0.3$. (a) Real part of the field at the horizon and at $\scri$, each channel normalised separately
to its own maximum. (b) Modulus on a logarithmic scale. Before the scattered packet arrives, $\scri$ records the small
outgoing residue of the flat-space initial data, which merges into the leading edge of the scattered signal; its effect
on the analysis is quantified in Sec.~\ref{sec:method}.}
\end{figure}

We record the full polar profile of $\psi$ at the horizon and at $\scri$ every $0.1M$. With
$\tilde\psi(\omega)=\int\psi\,e^{i\omega\tau}\dd\tau$ and the asymptotic forms $\psi\to Z^\infty$ at $\scri$ and
$\psi\to r_+Z^H$ at the horizon, we project at each real frequency onto
$\Sw_{Lm}(\theta;a\omega)$~\cite{BHPToolkit} to obtain $Z^\infty_L(\omega)$ and $Z^H_L(\omega)$. Negative frequencies
give the $(|\omega|,-m)$ modes through $R_{L,-m,-\omega}=\bar R_{Lm\omega}$. The energy fluxes are~\cite{Sasaki:2003xr}
\begin{align}
E^\infty_L&=\frac{|Z^\infty_L|^2}{4\pi\omega^2},\\
E^H_L&=\frac{128\,\omega k(k^2+4\epsilon^2)(k^2+16\epsilon^2)(2Mr_+)^5}{|C_L|^2}\,\frac{|Z^H_L|^2}{2\pi},
\end{align}
with $\epsilon=\sqrt{M^2-a^2}/(4Mr_+)$, and the greybody factor is
\begin{equation}
\Gamma_L=\frac{\omega}{\omega-m\Omega_H}\,\frac{E^H_L}{E^H_L+E^\infty_L}.
\label{eq:Gtd}
\end{equation}
Since $E^H_L\propto k$, the factor $k=\omega-m\Omega_H$ cancels, and we evaluate Eq.~\eqref{eq:Gtd} with it cancelled
analytically; $\Gamma_L$ is finite and positive at the superradiant threshold. The quantity that changes sign there is
the signed absorptivity
\begin{equation}
\mathcal{A}_L\equiv1-|R_L|^2=\frac{E^H_L}{E^H_L+E^\infty_L}=\frac{k}{\omega}\,\Gamma_L ,
\label{eq:absorptivity}
\end{equation}
which is negative in the superradiant band. Eq.~\eqref{eq:Gtd} uses only flux ratios, so the normalisation of the
incident packet drops out. A typical evolution is shown
in Fig.~\ref{fig:td}: the packet reaches the black hole, part of it is reflected to $\scri$ and part is absorbed,
and both signals end in quasinormal ringing.

 The evolution measures the outgoing wave relative to the horizon wave: with the horizon amplitude
normalised to unity, $Z^\infty_L/Z^H_L=B^{\rm ref}_L\,e^{i\varphi_0(\omega)}$, where the phase $\varphi_0$ comes from
the coordinates relating the horizon and $\scri$ and, at fixed $m$, is the same for all $L$. The incident amplitude
$B^{\rm inc}_L$ is not measured. The records are not windowed: they are transformed up to the end of the evolution,
which is why the evolution time matters (below).

Two features of the method matter in practice. First, the evolution time must exceed several damping times of the
least-damped mode. At $\chi=0.9$, truncating the record at $150M$, $200M$ and $250M$ biases $\Gamma_2$ in the
superradiant band by $0.9$, $0.1$ and $4\times10^{-3}$, while $400M$ gives $\le8.5\times10^{-5}$. We use $T_{\rm ev}=250M$ for
$\chi\le0.7$, $400M$ for $\chi=0.9$ and $700M$ for $\chi=0.99$. Second, a subdominant spheroidal mode can be measured
accurately only where it carries more than about $10^{-6}$ of the signal power; otherwise leakage from the dominant
mode through the angular projection limits the accuracy. We therefore excite each mode with its own incident
multipole $\ell'=L$.

\section{Validation}
\label{sec:validation}

The frequency-domain reference uses homogeneous Teukolsky solutions from the Black Hole Perturbation
Toolkit~\cite{BHPToolkit}, with $B^{\rm inc}$ obtained from the Wronskian,
$\Delta^{-1}(\Phi^{\rm in}\Phi^{\rm up\prime}-\Phi^{\rm up}\Phi^{\rm in\prime})=2i\omega B^{\rm inc}$. Two independent radial
integrators in the Toolkit agree to $10^{-11}$, and the superradiant amplification of the $\ell=m=2$ mode approaches
the known maximum of about $138\%$ as $\chi\to1$~\cite{Teukolsky:1974yv,Brito:2015oca} ($55\%$ at $\chi=0.99$, $133\%$ at
$\chi=0.99999$). At $\chi=0$ the Teukolsky greybody factor agrees with the Regge--Wheeler
one~\cite{Regge:1957td} to $3.5\times10^{-6}$. An external check of the dominant mode is provided by the public
complex reflection amplitude of the Kerr $(2,2)$ mode released with Ref.~\cite{Rosato:2026apq} ($\chi=0.001$--$0.999$,
$M\omega=0.001$--$1.6$, both in steps of $10^{-3}$), which uses the definition~\eqref{eq:RG} with
${\rm Im}\,C=+12M\omega$. Its phase differs from ours by exactly $4\ln2\,M\omega$ at every spin, i.e.\ by a different
additive constant in $r_*$. After removing this offset, the two complex amplitudes agree to
$|R_2/R_2^{\text{Ref.~\cite{Rosato:2026apq}}}-1|\le5\times10^{-5}$ for $\chi=0.1$, $8.5\times10^{-5}$ for $\chi=0.5$,
$1.3\times10^{-4}$ for $\chi=0.8$ and $4.5\times10^{-4}$ for $\chi=0.99$ over $0.1\le M\omega\le1.5$ (sampled every
$0.01$); a linear fit of the phase difference gives the slope $4\ln2$ to $6\times10^{-6}$; no complex conjugation
is involved, so the sign conventions agree for this $\ell+m$ even mode. Other recent FD
studies~\cite{Rosato:2025ulx,Huang:2025rxx} present greybody factors graphically, and the exact methods of
Ref.~\cite{Bonelli:2021uvf} provide another route.

\begin{table}[t]
\caption{\label{tab:val}Agreement between TD and FD greybody factors: the maximum relative difference
$\max|\Gamma^{\rm TD}_L/\Gamma^{\rm FD}_L-1|$ over all discrete-Fourier frequencies (spacing $1.9\times10^{-3}/M$) in
$0.3\le|M\omega|\le0.95$, both signs of $\omega$, for mode $L$ excited by incident multipole $\ell'=L$. The grid
includes the frequencies closest to the superradiant threshold, at $|k|=5.5\times10^{-5}$, $2.3\times10^{-4}$ and
$9.1\times10^{-4}$ for $\chi=0.7$, $0.9$ and $0.99$. The last column restricts to $|\Gamma_L|>10^{-3}$. $T_{\rm ev}$ is
the evolution time. At $\chi=0$ the radial equation does not depend on $m$. $^{\rm a}$Compared with the Regge--Wheeler
greybody factor over $0.25\le M\omega\le0.8$.}
\begin{ruledtabular}
\begin{tabular}{cccccc}
$\chi$ & $m$ & $L$ & $T_{\rm ev}/M$ & all $\omega$ & $|\Gamma_L|>10^{-3}$\\
\hline
0$^{\rm a}$ & -- & 2 & 250 & $2\times10^{-5}$   & $2\times10^{-5}$\\
0.3  & 2 & 2 & 250 & $1.8\times10^{-5}$ & $1.8\times10^{-5}$\\
0.7  & 1 & 2 & 250 & $2.0\times10^{-5}$ & $2.0\times10^{-5}$\\
0.7  & 2 & 2 & 250 & $3.9\times10^{-5}$ & $3.9\times10^{-5}$\\
0.7  & 2 & 3 & 250 & $1.3\times10^{-2}$ & $4.6\times10^{-3}$\\
0.7  & 2 & 4 & 250 & $2.3\times10^{-1}$ & $2.0\times10^{-3}$\\
0.9  & 2 & 2 & 400 & $8.5\times10^{-5}$ & $8.5\times10^{-5}$\\
0.9  & 2 & 3 & 400 & $3.4\times10^{-2}$ & $7.8\times10^{-3}$\\
0.99 & 2 & 2 & 700 & $1.3\times10^{-4}$ & $1.3\times10^{-4}$\\
\end{tabular}
\end{ruledtabular}
\end{table}

\begin{figure}[t]
\includegraphics[width=\columnwidth]{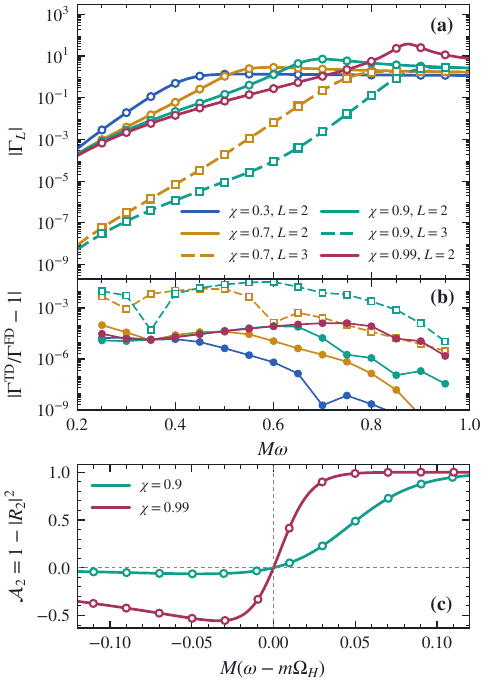}
\caption{\label{fig:val}(a) Per-mode Kerr greybody factors from 2+1 TD scattering (markers) and from the FD
(lines), for $\omega>0$; the negative frequencies (retrograde modes) are included in Table~\ref{tab:val}. For
$\chi=0.99$ the peak near $M\omega\simeq0.87$ is the superradiant threshold $m\Omega_H$, where $\Gamma_L$ is finite but
enhanced by the factor $\omega/(\omega-m\Omega_H)$ in Eq.~\eqref{eq:RG}. (b) Relative difference; filled markers
$L=2$, open markers $L=3$. (c) Signed absorptivity $\mathcal{A}_2=1-|R_2|^2=(k/\omega)\Gamma_2$ around the threshold,
for $\chi=0.9$ and $0.99$ (lines FD, markers TD): it changes sign at $k=0$.}
\end{figure}

Table~\ref{tab:val} and Fig.~\ref{fig:val} summarise the comparison. For the dominant mode the agreement is at the
$10^{-5}$ level for $\chi\le0.7$, for $m=2$ and also for $m=1$ (third row), and $\le1.3\times10^{-4}$ up to $\chi=0.99$,
including the superradiant band and the frequencies closest to the threshold $\omega=m\Omega_H$. Subdominant modes agree to better than $10^{-2}$ wherever $|\Gamma_L|>10^{-3}$;
deep in the tunnelling regime their small horizon signal is limited by the angular projection.
Fluxes and greybody factors are invariant under a common constant rescaling of the field. They are not invariant
under a change of the horizon amplitude alone, because Eq.~\eqref{eq:Gtd} combines the horizon and $\scri$ amplitudes
through separately normalised flux formulas; the agreement in Table~\ref{tab:val} therefore also tests the relative
normalisation of $E^H_L$ and $E^\infty_L$. We confirmed the absolute normalisation of the horizon flux separately: fitting a free
overall constant $n_H$ in place of $1/(2\pi)$ in $E^H$ at $\chi=0$ returns $n_H=0.159152$, against
$1/(2\pi)=0.159155$ (Appendix~\ref{app:norm}). Finally, the greybody factor of a given mode does not depend on the
incident packet: $\Gamma_3(M\omega=0.9)$ at $\chi=0.7$ is $1.8210$ when the mode is excited by leakage from an
$\ell'=2$ packet and $1.8207$ from an $\ell'=3$ packet, against the FD value $1.8198$.

The spherical-basis response~\eqref{eq:response} also needs the phases of $R_L$. Since $\varphi_0$ in
Sec.~\ref{sec:readout} is the same for every $L$, the ratio
\begin{equation}
\varrho_{L}=\frac{Z^\infty_L/Z^H_L}{B^{\rm ref}_L}\bigg/\frac{Z^\infty_2/Z^H_2}{B^{\rm ref}_2}
\label{eq:varrho}
\end{equation}
must equal unity. At $\chi=0.7$ we find $||\varrho_3|-1|\le6.4\times10^{-3}$ and $|\arg\varrho_3|\le5.7\times10^{-3}$ for
$0.3\le|M\omega|\le0.9$ (median $6\times10^{-4}$), and $||\varrho_4|-1|\le2\times10^{-2}$, $|\arg\varrho_4|\le1.3\times10^{-2}$ for
$|M\omega|\ge0.4$. At $\chi=0.9$, $||\varrho_3|-1|\le1.8\times10^{-2}$ and $|\arg\varrho_3|\le1.2\times10^{-2}$. The
evolution therefore reproduces the relative phases of the horizon-normalised reflected amplitudes $B^{\rm ref}_L$. The
relative phase of the energy-normalised amplitudes~\eqref{eq:RG} contains two further terms,
\begin{equation}
\arg\frac{R_L}{R_2}=\arg\frac{B^{\rm ref}_L}{B^{\rm ref}_2}-\arg\frac{B^{\rm inc}_L}{B^{\rm inc}_2}+\arg\frac{C_L}{C_2},
\label{eq:phasesplit}
\end{equation}
of which Eq.~\eqref{eq:varrho} tests only the first: $B^{\rm inc}_L$ comes from the Wronskian of the FD solutions,
and the phase of $C_L$ is the subject of Sec.~\ref{sec:parity}.

A coherent two-multipole packet tests the second term and the interference of Sec.~\ref{sec:rmatrix} more directly. At
$\chi=0.7$ we sent in the sum of the $\ell'=2$ and $\ell'=3$ packets of Eq.~\eqref{eq:indata} with equal weights; the
$\ell'=3$ packet carries $F'$ where the $\ell'=2$ one carries $F$, so its incident spectrum differs by a factor
$-i\omega$. The evolution is linear to $4\times10^{-9}$ (the combined run against the sum of the single-multipole runs).
The ratio of the spherical $\ell$ multipole at $\scri$ to that of the pure $\ell'=2$ run is predicted, without free
parameters, from the FD ratios $B^{\rm ref}_L/B^{\rm inc}_L$ and the mixing coefficients ($L\le5$), all
$\ell$-independent conventions cancelling. For $\ell=2$ the prediction agrees to $0.3$--$0.8\%$ in the complex ratio
for $M\omega=0.3$--$0.7$ ($1.6\%$ at $0.75$), and for $\ell=3$, where the $\ell'=3$ packet dominates, to $3$--$4\%$ for
$M\omega\le0.5$; with the opposite sign of the factor $-i\omega$ the prediction fails by $3$--$38\%$ and by factors of
$2$--$7$, respectively. At higher frequency the agreement degrades, to $10$--$30\%$, where one channel of the pure
$\ell'=2$ run is weak and the $\mathcal{O}(M/r_0)$ error of the flat-space incident data is comparable to the signal
being tested; a calibrated extraction of the incident wave would be needed to sharpen this test.

\section{Parity and the phase of the Teukolsky--Starobinsky constant}
\label{sec:parity}

Equation~\eqref{eq:C2} fixes $|C_L|$ but not its phase. Writing $C_L=D_L\pm12iM\omega$ with
$D_L=\sqrt{|C_L|^2-144M^2\omega^2}>0$, the two signs give reflection amplitudes~\eqref{eq:RG} with the same modulus
but phases that differ by $2\arctan(12M\omega/D_L)$. The greybody factor does not depend on the sign, but the phase of
$R_L$, and therefore the interference between modes in Eq.~\eqref{eq:response}, does. Ref.~\cite{Rosato:2025ulx}
works with the reflection probability $|R_L|^2$, in which the sign does not enter; Ref.~\cite{Rosato:2026apq} sets
${\rm Im}\,C=+12M\omega$ for the $(2,2)$, $(3,3)$ and $(4,4)$ modes, all with $\ell+m$ even.

At $\chi=0$ the reflection amplitudes of each parity can be computed directly from the Regge--Wheeler (axial) and
Zerilli (polar) equations~\cite{Regge:1957td,Zerilli:1970se}, which are related to the Teukolsky equation by the
Chandrasekhar transformations~\cite{Chandrasekhar:1975nkd,Chandrasekhar:1985kt}, which map the Teukolsky radial
function to the Regge--Wheeler and Zerilli functions.
We integrate both from the horizon with unit transmission and match to tenth-order asymptotic series at large $r$,
using the same $r_*$ as the Teukolsky solutions (Appendix~\ref{app:rwz}). The two equations are isospectral, as is
well known~\cite{Chandrasekhar:1975nkd,Chandrasekhar:1985kt}; numerically $|R_{\rm RW}|^2=|R_{\rm Z}|^2$ to
$3\times10^{-7}$, and their phases differ by $2\arctan(12M\omega/D)$, with $D=(\ell-1)\ell(\ell+1)(\ell+2)$: for $\ell=2$
we measure $\arg(R_{\rm Z}/R_{\rm RW})=0.2978$, $0.4900$ and $0.6731$ at $M\omega=0.3$, $0.5$ and $0.7$, against
$0.2978$, $0.4900$ and $0.6734$ from the formula.
Comparing with Eq.~\eqref{eq:RG}, we find
\begin{equation}
R_L\big|_{C=D+12iM\omega}=R_{\rm Z},\qquad R_L\big|_{C=D-12iM\omega}=R_{\rm RW},
\label{eq:parity}
\end{equation}
to $\le4.1\times10^{-4}$ for $\ell=2$ and $\le2.9\times10^{-5}$ for $\ell=3$ at $M\omega=0.3$, $0.5$ and $0.7$ (at
$\chi=0$ the radial equation does not depend on $m$). The opposite assignment fails at the level of the phase
difference $\delta\varphi$ itself ($|e^{i\delta\varphi}-1|\simeq0.3$--$0.7$ for $\ell=2$).

\begin{figure*}[t]
\includegraphics[width=\textwidth]{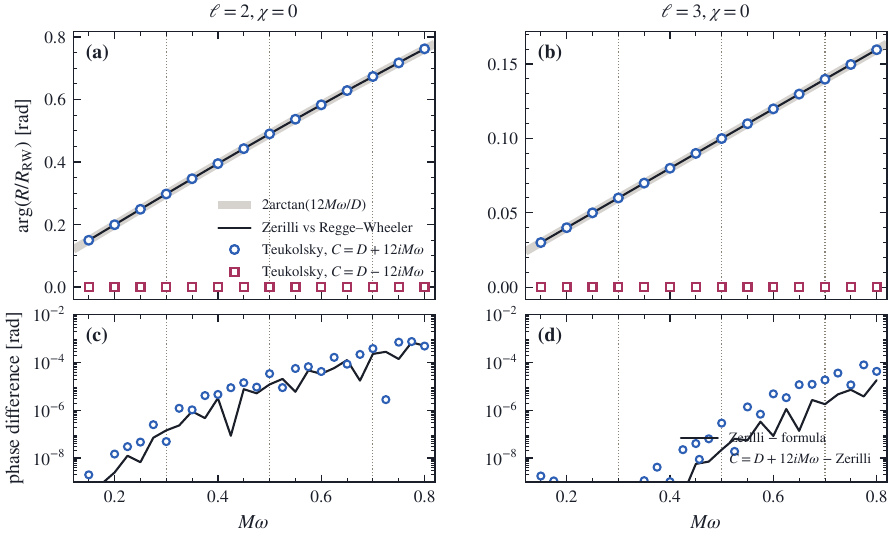}
\caption{\label{fig:parity}Phase of the energy-normalised reflection amplitude relative to the Regge--Wheeler one at
$\chi=0$, for (a) $\ell=2$ and (b) $\ell=3$. The thin line is the Zerilli amplitude, computed independently from the
Zerilli equation; the wide grey band is the Chandrasekhar relation $2\arctan(12M\omega/D)$. Teukolsky amplitudes with
$C=D+12iM\omega$ (circles) reproduce the Zerilli amplitude, and those with $C=D-12iM\omega$ (squares) the
Regge--Wheeler amplitude. (c), (d) Residuals: Zerilli minus the formula (line) and the Teukolsky amplitude with
$C=D+12iM\omega$ minus Zerilli (circles). Dotted vertical lines mark $M\omega=0.3$, $0.5$ and $0.7$, where
Eq.~\eqref{eq:parity} was tested.}
\end{figure*}

Figure~\ref{fig:parity} shows the comparison for $m=2$.

The sign of ${\rm Im}\,C$ therefore labels the parity of the reflected wave. This is not new in substance: it
follows from the Chandrasekhar transformations~\cite{Chandrasekhar:1975nkd,Chandrasekhar:1985kt} and underlies the
parity-dependent phase shifts used for the scattering of gravitational plane waves by Kerr black
holes~\cite{Futterman:1988ni,Dolan:2008kf}, where the combination $({\rm Re}\,C+12iM\omega P)B^{\rm ref}/B^{\rm inc}$
appears with $P=\pm1$; see also Ref.~\cite{Casals:2020fsb} for the properties of the Teukolsky--Starobinsky constants.
Here we state its consequence for greybody models of NR multipoles. For non-precessing binaries in the
Schwarzschild limit, the $\ell+m$ even multipoles [(2,2), (3,3), (4,4)] are polar, which corresponds to
$C=D+12iM\omega$, as used in Ref.~\cite{Rosato:2026apq}, while the $\ell+m$ odd multipoles [(2,1), (3,2), (4,3)] are
axial and correspond to $C=D-12iM\omega$. At $a\neq0$ a single spheroidal mode has no definite parity. We adopt
$C_L=D_L+(-1)^{L+m}\,12iM\omega$ at all spins as a prescription. Continuity labels the two branches unambiguously,
since $D_L>0$ means they never cross, but it does not by itself assign a branch to a merger source at finite spin;
that would require projecting the source onto the parity branches of the Kerr scattering
problem~\cite{Chandrasekhar:1985kt,Futterman:1988ni,Dolan:2008kf}, which we do not attempt. The NR data of
Sec.~\ref{sec:nr} cannot test it either.

The sign is observable only in restricted situations. If the sources $\sigma_L$ are free complex functions, the
replacement $R_L\to e^{i\delta_L}R_L$, $\sigma_L\to e^{-i\delta_L}\sigma_L$ leaves every multipole~\eqref{eq:response}
unchanged, for any number of modes, so the phase of $C_L$ is absorbed into the sources. It matters only when a model
fixes the relative phases of the sources, as in the spherical-basis description of Sec.~\ref{sec:rmatrix}, where the
sources are expanded in spherical multipoles.

\section{Mode mixing in the spherical basis}
\label{sec:mixing}

\begin{figure*}[t]
\includegraphics[width=\textwidth]{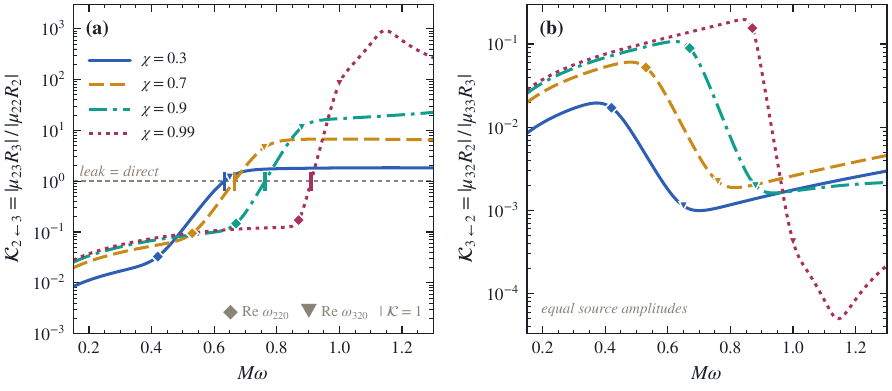}
\caption{\label{fig:kernels}Leakage kernels for $m=2$, for equal source amplitudes. (a) Contribution of the reflected
$L=3$ wave to the spherical $\ell=2$ channel relative to the direct $L=2$ reflection; the dashed line marks equality
and vertical bars the crossings $\mathcal{K}_{2\leftarrow3}=1$. For $\chi=0.99$ the curve is sampled every $0.002$ in
$M\omega$ near its maximum, about $920$ at $M\omega\simeq1.15$. (b) Contribution of the reflected $L=2$ wave to the
$\ell=3$ channel relative to the direct $L=3$ reflection (equal-source kernel; the leak in NR data is this kernel times
the source ratio). Diamonds mark ${\rm Re}\,\omega_{220}$ and triangles ${\rm Re}\,\omega_{320}$.}
\end{figure*}

Figure~\ref{fig:kernels} shows the leakage kernels for $m=2$ and $\chi=0.3$--$0.99$, computed from Eq.~\eqref{eq:response}
with the validated reflection amplitudes. Their behaviour follows from the barrier structure. Each $|R_L|$ is close to
unity below the peak of its own barrier and falls off exponentially above it; the $L=3$ barrier is higher, so
$|R_3/R_2|$ rises steeply across the window between ${\rm Re}\,\omega_{220}$ and ${\rm Re}\,\omega_{320}$ and then
saturates above it, once both amplitudes fall off. The mixing coefficients $\mu_{23}$ and $\mu_{32}$ grow linearly
with $a\omega$. As a result:

\emph{The $\ell=2$ channel.} Below ${\rm Re}\,\omega_{220}$ the leak is negligible. Above it, the reflected $L=3$
wave overtakes the direct one, $\mathcal{K}_{2\leftarrow3}=1$, at $M\omega=0.632$, $0.664$, $0.762$ and $0.910$ for
$\chi=0.3$, $0.7$, $0.9$ and $0.99$. At $M\omega=1$, near the upper edge of the band usable in the NR data of
Sec.~\ref{sec:nr}, $\mathcal{K}_{2\leftarrow3}=1.8$, $6.8$, $17$ and $84$; at $M\omega=1.2$ it is $1.8$, $6.6$, $20$ and
$590$, and for $\chi=0.99$ it peaks at about $920$ near $M\omega=1.15$. These values hold for equal source amplitudes;
the leak in an NR multipole is $\mathcal{K}$ times the source ratio, which we infer in Sec.~\ref{sec:nr}. The growth
with spin reflects the larger mixing coefficients and the steeper fall-off of $R_2$ above the superradiant threshold.
The kernels for $m=1$, $3$ and $4$ are given in Appendix~\ref{app:otherm}.

\emph{The $\ell=3$ channel.} Below ${\rm Re}\,\omega_{220}$ both modes are fully reflected and
$\mathcal{K}_{3\leftarrow2}\simeq|\mu_{32}/\mu_{33}|$, between about $1\%$ and $20\%$ in this spin range. The
$(2,2)$ source is typically much stronger than the $(3,2)$ source, so this is where the leak matters most in practice;
whether it dominates the $(3,2)$ multipole depends on the source ratio (Sec.~\ref{sec:nr}).

\begin{figure*}[t]
\includegraphics[width=\textwidth]{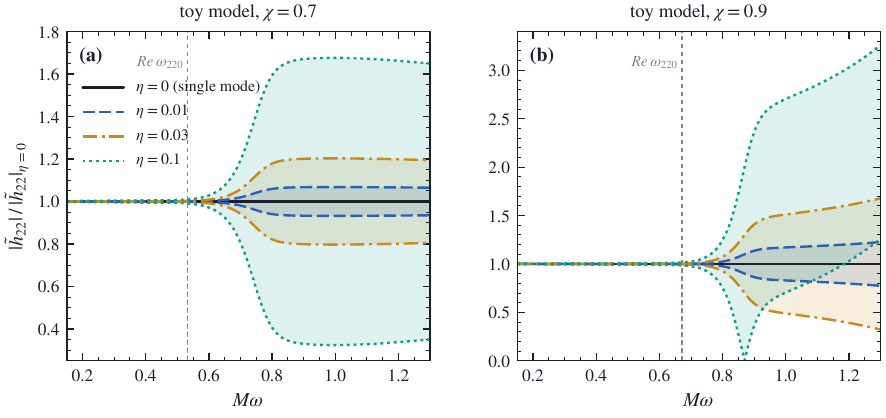}
\caption{\label{fig:toy}Toy model of the spherical $(2,2)$ spectrum including the leaked $L=3$ reflection, for
frequency-independent sources with $|\sigma_3/\sigma_2|=\eta$ and a frequency-independent relative phase,
normalised to the single-mode spectrum. Bands cover all relative phases; their edges are the analytic envelope
$|1-\eta\mathcal{K}_{2\leftarrow3}|$ and $1+\eta\mathcal{K}_{2\leftarrow3}$. Dashed vertical line:
${\rm Re}\,\omega_{220}$. The two panels use different vertical scales.}
\end{figure*}

To illustrate the size of the effect on the dominant multipole, Fig.~\ref{fig:toy} shows
$\tilde h_{22}=\mu_{22}R_2\sigma_2+\mu_{23}R_3\sigma_3$ in a toy model with frequency-independent source amplitudes,
$|\sigma_3/\sigma_2|=\eta$, and a frequency-independent relative phase that we vary over all values. The ratio to the
single-mode spectrum then lies between $|1-\eta\mathcal{K}_{2\leftarrow3}|$ and $1+\eta\mathcal{K}_{2\leftarrow3}$.
For $\eta=0.03$ the spectral amplitude at $M\omega=1.2$ changes by up to $20\%$ at $\chi=0.7$ and $61\%$ at $\chi=0.9$
(and by a factor of $19$ at $\chi=0.99$, not shown). Because within the fit window below the leaked term grows smoothly
relative to the direct one, a single-mode fit with a free exponential factor absorbs part of it: fitting $\ln|\tilde h_{22}|^2=c_{\rm fit}-\omega/T_{\rm eff}$ over
${\rm Re}\,\omega_{220}\le M\omega\le\min(1.3,1.6\,{\rm Re}\,\omega_{220})$, the toy-model $T_{\rm eff}$ shifts relative to the
pure-$R_2$ value by $\pm4\%$ at $\chi=0.7$ and by $+24$--$28\%$ at $\chi=0.99$, depending on the relative phase.

\subsection{Spherical-basis reflection matrix}
\label{sec:rmatrix}
Equation~\eqref{eq:response} can be written compactly as a matrix acting on the source. Expand the source in
spherical multipoles, $\sigma_L=\sum_{j}\mu_{jL}s_{j}$, where $s_j$ is the amplitude that the source would radiate into
the spherical multipole $j$ if the barrier were absent ($R_L=1$ for all $L$). This is the natural description of
sources computed far from the black hole, e.g.\ in post-Newtonian or effective-one-body waveforms, which are
expanded in spherical harmonics. Then
\begin{align}
\tilde h_{\ell m}&=\sum_{j}\mathcal{R}_{\ell j}\,s_{j},\nonumber\\
\mathcal{R}_{\ell j}(\omega)&=\sum_L\mu_{\ell L}(a\omega)\,R_L(\omega)\,\mu_{jL}(a\omega).
\label{eq:Rmatrix}
\end{align}
At $a=0$ this reduces to the single-mode model $\tilde h_{\ell m}=R_\ell s_\ell$. Since the $\mu_{\ell L}$ form a real
orthogonal matrix at real frequency, $\mathcal{R}$ is symmetric and the energy radiated by a pure source $s_j$ obeys
\begin{align}
\sum_\ell|\mathcal{R}_{\ell j}|^2&=1-\frac{\omega-m\Omega_H}{\omega}\,\Gamma^{\rm eff}_{j},\nonumber\\
\Gamma^{\rm eff}_{j}&=\sum_L\mu_{jL}^2\,\Gamma_L .
\label{eq:Geff}
\end{align}
Both properties follow directly from the orthogonality of $\mu$; the symmetry holds by construction, while
Eq.~\eqref{eq:Geff} provides a numerical check. Computed for $m=2$ with spheroidal modes $L\le6$ and the parity rule
of Sec.~\ref{sec:parity}, Eq.~\eqref{eq:Geff} holds to $2.5\times10^{-5}$, the level of the truncation in $L$; the
$\Gamma_L$ are computed from the horizon flux, Eqs.~\eqref{eq:Gtd}--\eqref{eq:absorptivity}, rather than from
$1-|R_L|^2$, which loses precision where $\Gamma_L\lesssim10^{-8}$.

\begin{figure*}[t]
\includegraphics[width=\textwidth]{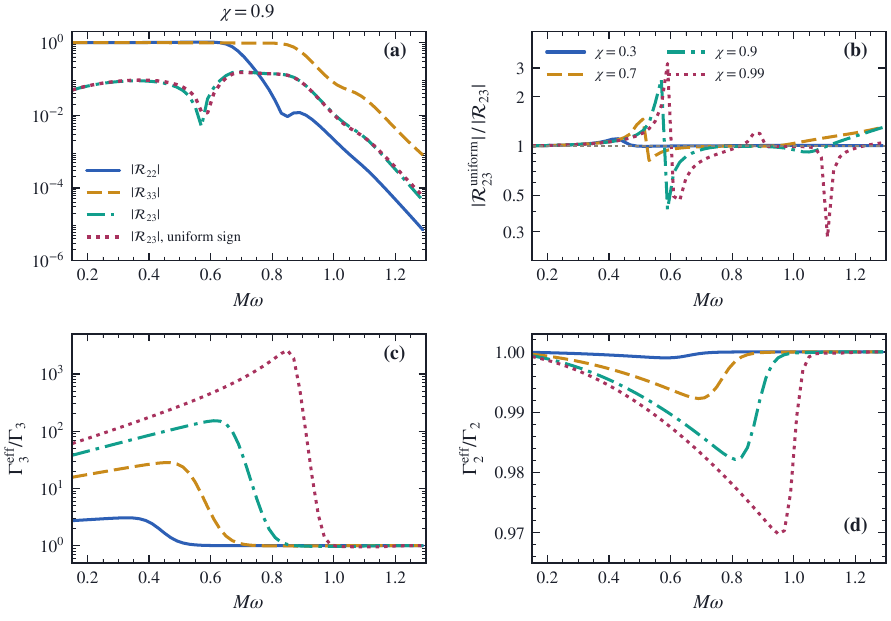}
\caption{\label{fig:rmatrix}Spherical-basis reflection matrix, Eq.~\eqref{eq:Rmatrix}, for $m=2$, computed with
spheroidal modes $L\le6$. (a) Entries for $\chi=0.9$; the dotted line shows $|\mathcal{R}_{23}|$ with a uniform sign
$C_L=D_L+12iM\omega$ instead of the parity-resolved one. (b) Ratio of the two $|\mathcal{R}_{23}|$ for four spins.
(c) Effective greybody factor of a spherical $(3,2)$ source relative to the spheroidal one, Eq.~\eqref{eq:Geff}, with
$\Gamma_L$ from the horizon flux. (d) The same for a $(2,2)$ source.}
\end{figure*}

Figure~\ref{fig:rmatrix} shows the features that matter in practice. For $\chi=0.9$ the off-diagonal entry reaches
$|\mathcal{R}_{23}|\simeq0.15$ at ${\rm Re}\,\omega_{220}$ and exceeds the diagonal entry $|\mathcal{R}_{22}|$ above
$M\omega=0.75$, as anticipated by the leakage kernels. The off-diagonal entries are where the parity rule enters:
replacing the parity-resolved Teukolsky--Starobinsky phases by a uniform $+12iM\omega$ changes the complex
$\mathcal{R}_{23}$ by a median of $15$--$24\%$ over the four spins; its modulus changes by a median of only $0.7$--$5\%$,
but at some frequencies by factors between $0.3$ and $3.2$ for $\chi\ge0.9$. The effective greybody factor of a spherical
$(2,2)$ source differs from $\Gamma_2$ by at most $3\%$. For a spherical $(3,2)$ source below its own barrier, the
absorbed fraction is set by its small $L=2$ component, $\Gamma^{\rm eff}_3\simeq\mu_{32}^2\Gamma_2$, which tunnels far
more easily; $\Gamma^{\rm eff}_3/\Gamma_3$ reaches $3$, $28$, $150$ and $2600$ for $\chi=0.3$, $0.7$, $0.9$ and $0.99$. A
greybody model that assigns $\Gamma_3$ to a source computed in the spherical basis therefore underestimates its
absorption by these factors.

\section{Comparison with numerical relativity}
\label{sec:nr}

\subsection{Data}
We use nine non-precessing simulations from the SXS catalog~\cite{Boyle:2019kee,Scheel:2025jct}. The first set is the
three near-equal-mass mergers analysed with greybody factors in Ref.~\cite{Okabayashi:2024qbz}: SXS:BBH:0305, 0540 and
1124, with mass ratios $q=1.22$, $1.09$, $1.00$ and remnant spins $\chi_f=0.692$, $0.832$, $0.951$. The second set
consists of unequal-mass mergers with aligned spins and high remnant spin: SXS:BBH:4121, 3128 and 2786, with $q=4.0$,
$4.5$, $8.0$ and $\chi_f=0.928$, $0.874$, $0.862$. The third set consists of unequal-mass mergers in which the
secondary spin is anti-aligned, so that the remnant spin is moderate: SXS:BBH:0317, 1443 and 2669, with $q=3.3$, $5.7$,
$8.0$ and $\chi_f=0.746$, $0.631$, $0.585$; for SXS:BBH:1443 and 2669 the inferred $(3,2)$ source is the strongest of
all nine runs relative to the $(2,2)$ one (Table~\ref{tab:nr}). The component spins are listed in
Table~\ref{tab:nr} and the run manifest in Appendix~\ref{app:manifest}. We use the extrapolated ($N=2$) strain of
catalog version 3.0 at the highest available resolution (Lev3--Lev6), in units of the remnant mass $M_f$. For the error
budget (Sec.~\ref{sec:robust}) we also use the $N=3$ and $N=4$ extrapolations of the six runs of the first two sets,
and the second-highest resolution of the seven runs for which it is available (all except SXS:BBH:1124 and 4121).

The records end $141$--$362M_f$ after the peak of $|h_{22}|$, when $|h_{22}|$ is still $2\times10^{-5}$--$4\times10^{-4}$
of its peak, so we apply smooth tapers at both ends; the start of the taper, at $-300M_f$, does not affect the post-peak
spectrum (Sec.~\ref{sec:robust}). Equatorial symmetry relates the $(\ell,m,\omega)$ and $(\ell,-m,-\omega)$ content but
does not forbid counter-rotating emission at $\omega<0$ in an $m=2$ multipole. We use $|\tilde h_{\ell2}(-\omega)|$ only as
an empirical estimate of the floor set by numerical noise and by the tapers, and restrict the analysis to frequencies
where $|\tilde h_{\ell2}(\omega)|$ exceeds it by a factor of $30$, up to the first frequency where this fails. This gives
upper limits $M_f\omega\simeq1.0$--$1.3$. The lower edge of the band is $M_f\omega=0.3$. This floor estimate does not
capture all errors: leakage from the end taper of the positive-frequency signal need not be symmetric in $\omega$, and
numerical truncation error is not included. We return to this in Secs.~\ref{sec:robust} and~\ref{sec:excess}.

\subsection{Mixing coefficients}
\label{sec:qnmratio}
As a first check we fit the post-peak time-domain multipoles with $\{\omega_{220},\omega_{221},\omega_{320}\}$
(optionally adding $\omega_{222}$) for $h_{22}$ and $\{\omega_{220},\omega_{221},\omega_{320},\omega_{321}\}$ for
$h_{32}$, using QNM frequencies from Ref.~\cite{Stein:2019mop}, with fit start times $10$--$20M_f$ after the peak and
windows of $40$ and $60M_f$. The complex ratio of the $(2,2,0)$ amplitudes in $h_{32}$ and $h_{22}$ should equal
$\mu_{32}/\mu_{22}$ evaluated at the complex QNM frequency, which we take from the spherical--spheroidal expansion of
Ref.~\cite{Stein:2019mop}. In eight of the nine runs the range of measured moduli over fit choices contains the
prediction (Table~\ref{tab:nr}); for SXS:BBH:0305 it lies $1$--$4\%$ below it. The spread over fit choices is
$2\%$--$15\%$, largest at the highest spins. For the phases, the prediction lies inside the fit range for the four runs
with $\chi_f\ge0.86$, where the range is wide, while in the other five the measured phase is less negative than the
prediction by $0.03$--$0.13$\,rad. These offsets are at the level at which the asymptotic frame of the NR
waveform~\cite{Boyle:2015nqa,Mitman:2022kwt,MaganaZertuche:2021syq} and the remnant parameters affect multimode
fits~\cite{Cook:2020otn,MaganaZertuche:2021syq}. This confirms the coefficients and the harmonic conventions used
below. The same QNM mixing has been studied in detail in the time
domain~\cite{Kelly:2012nd,London:2014cma,Berti:2014fga,Cook:2020otn,MaganaZertuche:2021syq}; here it serves to validate
the FD treatment.

\begin{table*}[t]
\caption{\label{tab:nr}NR tests of the mixing. Rows 3--4: component spins. Rows 5--8: complex $(2,2,0)$ amplitude in
$h_{32}$ relative to $h_{22}$ from time-domain QNM fits (range over fit start, window and overtone content), against
$\mu_{32}/\mu_{22}$ at the complex QNM frequency. Row 9: median ratio of the inferred sources. Row 10: median ratio of
the leaked $(2,2)$ reflection to the direct $(3,2)$ reflection in $\tilde h_{32}$ over $[0.3,\omega_{\rm top}]$.
Row 11: mixing strength preferred by the data (complex scan, baseline analysis). Rows 12--13: range of the preferred
$\kappa$ over the baseline and nine variants of Sec.~\ref{sec:robust}. Row 14: preferred $|\kappa|$ for the $N=3$ and
$N=4$ extrapolations and the second-highest resolution where available; for SXS:BBH:1124 and 4121 only $N=3,4$, for
the third set only the second resolution. The run manifest is given in Appendix~\ref{app:manifest}.}
\begin{ruledtabular}
\footnotesize
\setlength{\tabcolsep}{1.6pt}
\begin{tabular}{lccccccccc}
 & 0305 & 0540 & 1124 & 4121 & 3128 & 2786 & 0317 & 1443 & 2669\\
\hline
$q$ & 1.22 & 1.09 & 1.00 & 4.00 & 4.50 & 8.00 & 3.33 & 5.68 & 8.00\\
$\chi_f$ & 0.692 & 0.832 & 0.951 & 0.928 & 0.874 & 0.862 & 0.746 & 0.631 & 0.585\\
$\chi_{1z}$ & $+0.33$ & $+0.36$ & $+0.998$ & $+0.90$ & $+0.80$ & $+0.80$ & $+0.52$ & $+0.41$ & $+0.40$\\
$\chi_{2z}$ & $-0.44$ & $+0.69$ & $+0.998$ & $+0.50$ & $0.00$ & $+0.40$ & $-0.45$ & $-0.74$ & $-0.80$\\
$|A_{220}^{(32)}/A_{220}^{(22)}|$ & $0.067$--$0.069$ & $0.093$--$0.097$ & $0.130$--$0.150$ & $0.119$--$0.131$ & $0.103$--$0.110$ & $0.100$--$0.107$ & $0.077$--$0.079$ & $0.059$--$0.062$ & $0.052$--$0.056$\\
$|\mu_{32}/\mu_{22}|$ & $0.070$ & $0.096$ & $0.134$ & $0.124$ & $0.107$ & $0.104$ & $0.079$ & $0.061$ & $0.055$\\
$\arg(A_{220}^{(32)}/A_{220}^{(22)})$ & $-0.09$--$-0.05$ & $-0.08$--$0.01$ & $-0.21$--$0.27$ & $-0.19$--$0.14$ & $-0.13$--$0.05$ & $-0.12$--$0.06$ & $-0.11$--$-0.06$ & $-0.13$--$-0.09$ & $-0.14$--$-0.09$\\
$\arg(\mu_{32}/\mu_{22})$ & $-0.15$ & $-0.12$ & $-0.07$ & $-0.08$ & $-0.11$ & $-0.11$ & $-0.14$ & $-0.16$ & $-0.17$\\
$|\sigma_3/\sigma_2|$ & $0.028$ & $0.011$ & $0.010$ & $0.005$ & $0.010$ & $0.022$ & $0.006$ & $0.030$ & $0.042$\\
leak/direct in $\tilde h_{32}$ & $0.12$ & $0.89$ & $7.9$ & $5.0$ & $2.1$ & $1.7$ & $0.76$ & $0.11$ & $0.04$\\
best $\kappa$ & $1.04\,e^{0.01i}$ & $0.97\,e^{0.07i}$ & $0.97\,e^{0.12i}$ & $0.96\,e^{0.06i}$ & $0.94\,e^{0.08i}$ & $0.94\,e^{-0.04i}$ & $0.98\,e^{0.05i}$ & $0.91\,e^{0.08i}$ & $0.82\,e^{0.13i}$\\
$|\kappa|$, variants & $0.96$--$1.06$ & $0.96$--$1.00$ & $0.97$ & $0.96$--$0.98$ & $0.93$--$1.15$ & $0.93$--$1.30$ & $0.97$--$0.99$ & $0.90$--$0.97$ & $0.82$--$1.01$\\
$\arg\kappa$, variants & $0.00$--$0.11$ & $-0.02$--$0.09$ & $0.11$--$0.13$ & $0.04$--$0.08$ & $-0.04$--$0.13$ & $-0.14$--$0.00$ & $0.04$--$0.06$ & $0.06$--$0.13$ & $0.11$--$0.21$\\
$|\kappa|$, $N$ and Lev & $1.03$ & $0.97$--$0.98$ & $0.97$--$0.98$ & $0.97$ & $0.99$ & $0.95$--$1.13$ & $0.97$ & $0.90$ & $0.84$\\
\end{tabular}
\end{ruledtabular}
\end{table*}

\subsection{Decomposition of the spectra}
\begin{figure*}[t]
\includegraphics[width=\textwidth]{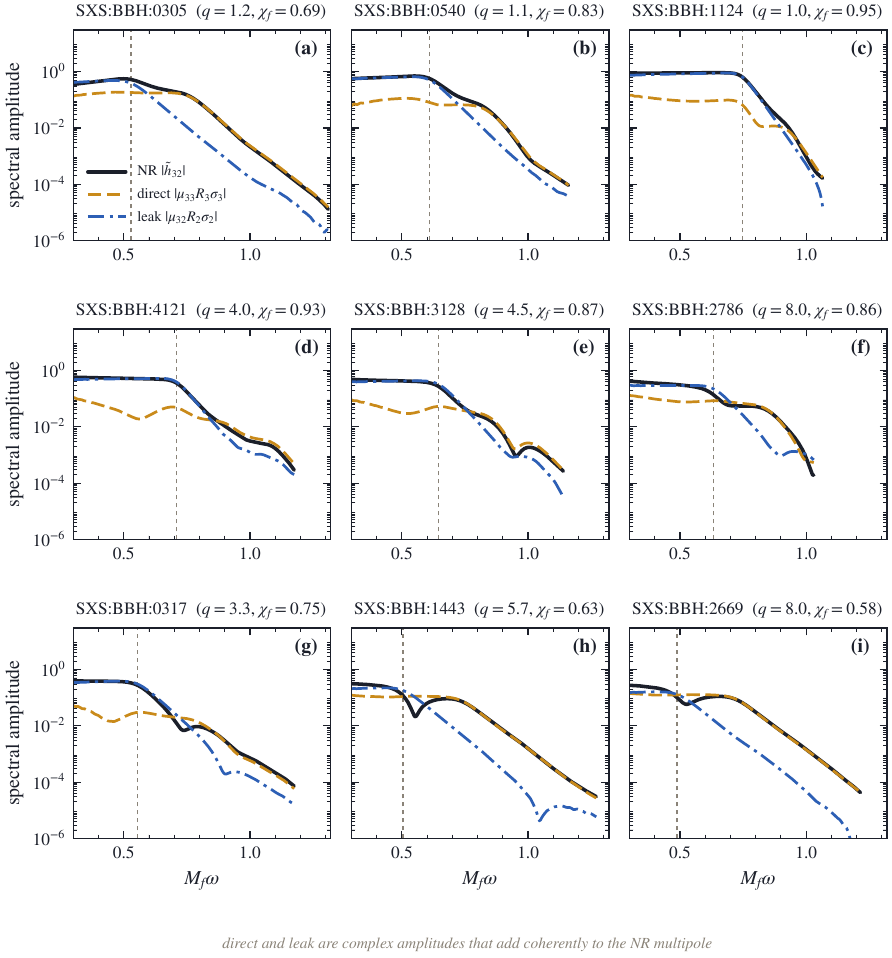}
\caption{\label{fig:nrdecomp}Spectral amplitude of the $(3,2)$ multipole of the nine SXS mergers (solid) and its
decomposition into the direct $L=3$ reflection and the leaked $L=2$ reflection, Eq.~\eqref{eq:response}; the two
complex terms add coherently to the data. Top row: near-equal-mass mergers; middle row: unequal-mass mergers with
$\chi_f\ge0.86$; bottom row: unequal-mass mergers with $\chi_f\le0.75$. Each panel ends where the signal reaches $30$ times the floor. Dashed vertical lines: ${\rm Re}\,\omega_{220}$.}
\end{figure*}

At each frequency we invert Eq.~\eqref{eq:response} for $\ell,L\in\{2,3\}$ to obtain the sources
$\sigma_2(\omega)$ and $\sigma_3(\omega)$ from $\tilde h_{22}$ and $\tilde h_{32}$. The inferred ratio
$|\sigma_3/\sigma_2|$ is small in all nine runs, $0.005$--$0.04$ (Table~\ref{tab:nr}); it is largest for the runs
with an anti-aligned secondary spin at large mass ratio and for the smallest remnant spin.
Figure~\ref{fig:nrdecomp} decomposes $\tilde h_{32}$ into its direct and leaked parts for all nine runs. In a square
inversion the two parts add coherently to reproduce $\tilde h_{32}$ exactly, so the decomposition is not a test of
either barrier; it shows how much of $\tilde h_{32}$ must be removed before a greybody model of the $(3,2)$ mode is
fitted. Below ${\rm Re}\,\omega_{220}$ the leaked $(2,2)$ reflection dominates the $(3,2)$ spectrum in the first six
runs, by median amplitude factors of $2.7$--$11$; in the three runs of the third set the two are comparable or the
direct reflection dominates. Above ${\rm Re}\,\omega_{220}$, where $R_2$ falls off, the direct reflection takes over.
Over the whole band the per-run median of the ratio of the leaked to the direct amplitude ranges from $0.04$ to $7.9$
(Table~\ref{tab:nr}). These are ratios of amplitudes that add coherently, not energy fractions; their effect on
$|\tilde h_{32}|$ depends on the relative phase. Because $\sigma_2$ is inferred from $\tilde h_{22}$, the leaked part
is essentially $(\mu_{32}/\mu_{22})\,\tilde h_{22}$, so it is not a test of $R_2$. A single-mode greybody description
of the $(3,2)$ multipole therefore fails at high spin. In $\tilde h_{22}$ the leaked $L=3$ reflection reaches, at the
top of the usable band, $8\%$ (SXS:BBH:4121) to $38\%$ (2786) of the direct one in the first six runs, and $10\%$, $66\%$
and $38\%$ in SXS:BBH:0317, 1443 and 2669.

\subsection{Data-preferred mixing strength}
\label{sec:kappa}

\begin{figure*}[t]
\includegraphics[width=\textwidth]{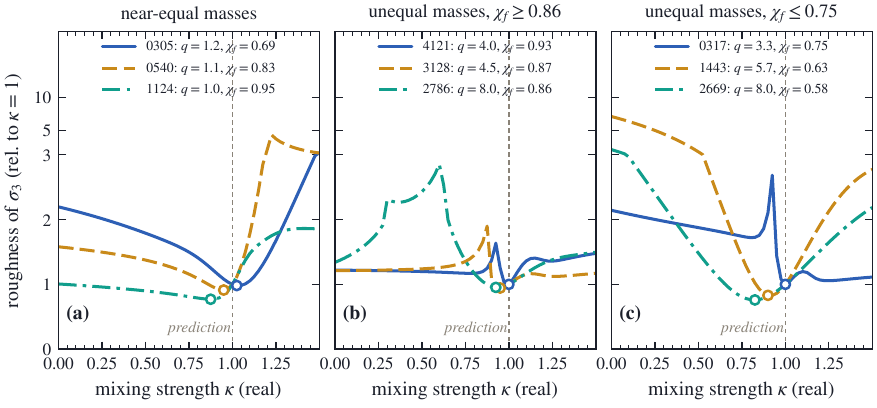}
\caption{\label{fig:kappa}Roughness of the inferred $(3,2)$ source as a function of a real strength $\kappa$ of the
off-diagonal mixing, normalised to its value at the prediction $\kappa=1$, for (a) the near-equal-mass mergers and the unequal-mass
mergers with (b) $\chi_f\ge0.86$ and (c) $\chi_f\le0.75$. Circles mark the minima of the real scan, which can differ
from the complex best $\kappa$ of Table~\ref{tab:nr} (Fig.~\ref{fig:kappacplx}); $\kappa=0$ is the single-mode description. The spikes in (b) occur where $\sigma_3$ passes close
to zero and its unwrapped phase jumps.}
\end{figure*}
To test the size of the mixing directly, we scale the off-diagonal coefficients, $\mu_{23}\to\kappa\mu_{23}$ and
$\mu_{32}\to\kappa\mu_{32}$, and measure how smooth the inferred source $\sigma_3(\omega)$ is. Our measure of roughness
is the rms residual of $\ln|\sigma_3|$ plus that of its unwrapped phase, about a cubic polynomial over the band. A
physical source produced by the plunge should be smooth in frequency, while a wrong mixing strength leaves imprints of
$R_2$ and $R_3$. The prediction is $\kappa=1$; $\kappa=0$ is the single-mode description. In the $2\times2$ inversion
$R_2$ cancels from $\sigma_3$ and $R_3$ is an overall factor, so this is a FD test of the angular mixing coefficients,
complementary to the QNM fits above; we test the reflection amplitude separately below.

Table~\ref{tab:scores} gives the roughness at fixed values of $\kappa$, relative to the best complex $\kappa$:
$\mathcal{S}_0$ for the single-mode description, $\mathcal{S}_1$ for the prediction, and $\mathcal{S}_{\rm real}$ for
the best real $\kappa$. Since the best complex $\kappa$ has two fitted parameters, the parameter-free comparison of
the prediction with the single-mode description is the ratio $\mathcal{S}_0/\mathcal{S}_1$ of the roughness at
$\kappa=0$ and $\kappa=1$. To judge these numbers we use two controls. In a null test we pair the $\tilde h_{32}$ of each
run with the $\tilde h_{22}$ of each of the other eight runs, which carries no information on the mixing of that run,
and apply the same band and roughness measure to all $72$ pairings. The null distribution of $\mathcal{S}_0$ has median
$1.02$ and 95th percentile $1.89$; that of $\mathcal{S}_0/\mathcal{S}_1$ has median $0.86$ and 95th percentile $1.64$,
and in $90\%$ of the pairings the best $\kappa$ lies at the edge of the scan. The null is not perfectly clean: all
$\tilde h_{22}$ share the shape imposed by $R_2$, so an unrelated $\tilde h_{22}$ still removes part of the leaked
$(2,2)$ reflection. The two largest null values of $\mathcal{S}_0$, $6.4$ and $3.3$, both use the $\tilde h_{32}$ of
SXS:BBH:1443, whose leak is strong. In a mock test we build $\tilde h_{22}$ and $\tilde h_{32}$ from
Eq.~\eqref{eq:response} with $\kappa=1$ and smooth sources (cubic fits to the inferred ones), add the measured
difference between the $N=2$ and $N=3$ extrapolations (for the third set, between the two resolutions) as noise, and
apply the same analysis. It recovers $|\kappa|=0.99$--$1.04$, with $\mathcal{S}_1\le1.11$. The mocks are, however, much
cleaner than the data ($\mathcal{S}_0=4.4$--$36$ against $1.4$--$10$; $35.7$ against $1.49$ for SXS:BBH:3128): they
show that the analysis recovers the input mixing strength, but they do not calibrate the significance of the NR
results.

\begin{figure}[!htb]
\includegraphics[width=\columnwidth]{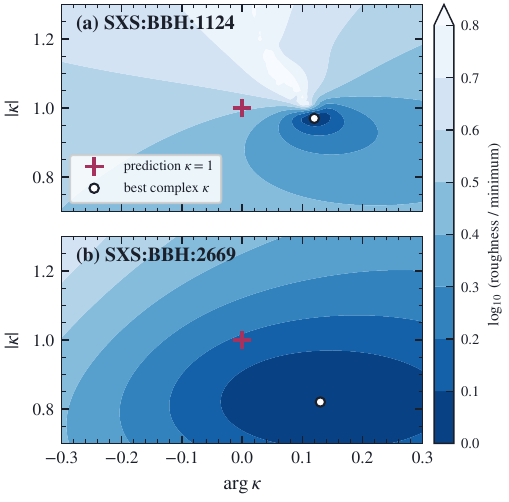}
\caption{\label{fig:kappacplx}Roughness of the inferred $(3,2)$ source over complex $\kappa$ near $\kappa=1$, in
$\log_{10}$ relative to the minimum, for (a) SXS:BBH:1124, where the phase of $\kappa$ matters most, and
(b) SXS:BBH:2669, where the best $|\kappa|$ is lowest ($0.82$). Cross: prediction $\kappa=1$; circle: best complex
$\kappa$.}
\end{figure}

With a fitted mixing strength the single-mode description is disfavoured beyond the 95th percentile of the null in
seven runs, with $\mathcal{S}_0=2.0$--$10$ (97th to 100th percentile), most strongly in the third set ($5.1$--$10$); for
SXS:BBH:3128 and 2786, $\mathcal{S}_0=1.49$ and $1.44$ lie at the 90th and 89th percentiles. With the strength fixed at the
prediction, $\mathcal{S}_0/\mathcal{S}_1=2.1$--$6.7$ exceeds the 95th percentile of the null in four runs (SXS:BBH:0305,
0317, 1443 and 2669); for SXS:BBH:0540 it is $1.58$ (94th percentile), for SXS:BBH:4121, 3128 and 2786 it is
$1.21$--$1.34$ (92nd--93rd), and for SXS:BBH:1124 it is $1.01$. The prediction $\kappa=1$ comes within a factor
$1.02$--$1.73$ of the best roughness in seven runs ($\mathcal{S}_1$), rougher than in the mocks, which indicates
structure in the NR sources or errors beyond the extrapolation differences. For SXS:BBH:1124 and 0317,
$\mathcal{S}_1=3.3$ and $2.4$: there the preferred phase of $\kappa$ matters, and for SXS:BBH:1124, where the leak is
about eight times the direct signal, $\kappa=1$ is as rough as $\kappa=0$. The complex scan, which adds two free
parameters and breaks the real orthogonality of $\mu$, is a diagnostic of this phase offset: it gives
$|\kappa|=0.91$--$1.04$ in eight runs and $0.82$ for SXS:BBH:2669, with phases between $-0.04$ and $0.13$
(Table~\ref{tab:nr}); Fig.~\ref{fig:kappacplx} shows two examples. Figure~\ref{fig:kappa} shows the real-$\kappa$ scans. In the unequal-mass runs of the second set
the roughness has sharp spikes at intermediate $\kappa$, where $\sigma_3$ passes close to zero and its phase unwrapping
jumps, but the minima again lie at $\kappa=0.94$--$0.98$.

The $\ell=2$ channel, analysed in the same way for $\sigma_2$, prefers $\kappa\simeq0.9$ for $\chi_f=0.83$ and $0.95$. In
the third set the leak into $\tilde h_{22}$ reaches $10\%$--$66\%$ of the direct reflection at the top of the baseline
band, so this channel should be most sensitive there, but the result is mixed: for SXS:BBH:1443 demixing makes $\sigma_2$ smoother by a factor of $1.9$ (rms
log-amplitude residual $0.031\to0.019$, phase $0.034\to0.016$), whereas for SXS:BBH:2669 the single-mode source is the
smoother one. The $(2,2)$ channel thus supports the mixing in SXS:BBH:0540, 1124 and 1443 but not in 2669; with these data it is consistent with the result from $\sigma_3$ but does not confirm it independently. In a control the mixing
coefficients are evaluated at the wrong-sign spin, $\mu(-a\omega)$, so the off-diagonal terms change sign and
approximately keep their size ($|\mu_{32}|=0.263$ at $c=1.5$ and $0.241$ at $c=-1.5$). For polynomial degrees $4$ and $5$
this makes the phase of the inferred $(2,2)$ source of the near-equal-mass runs rougher than with the physical
coefficients. The parity prescription of Sec.~\ref{sec:parity} cannot be tested this way, because $R_3$ cancels from
$\sigma_2$ and enters $\sigma_3$ only as an overall factor.

\emph{Sensitivity to the reflection amplitude.} The greybody hypothesis proper is that the demixed spectrum divided by
$R_3$ is smooth. At fixed mixing ($\kappa=1$) we form the demixed spheroidal component $b_3=(\mu^{-1}\tilde h)_3$ and
divide it once by each trial amplitude: $R_3$, no barrier ($R\to1$), the wrong mode ($R_2$), and $R_3$ at
$\chi_f\pm0.05$ (Table~\ref{tab:scores}, roughness relative to $R_3$). For $\chi_f=0.69$ and $0.83$ the nominal $R_3$
gives a smoother source than no barrier, by factors of $3.8$ and $2.2$, and than $R_2$, by $5.4$ and $3.5$. The same
holds for the three runs of the third set ($\chi_f=0.59$--$0.75$): no barrier is rougher by $1.4$--$4.5$ and $R_2$ by
$1.7$--$6.6$. The data do not resolve the remnant spin in $R_3$, however: $R_3$ at $\chi_f-0.05$ is as smooth or
smoother ($0.56$--$1.06$) in four of these five runs. For the four runs with $\chi_f\ge0.86$ all alternatives are
equally smooth, within factors $0.8$--$1.4$: in these runs the leak exceeds the direct $(3,2)$ signal by factors of
$1.7$--$8$ (Table~\ref{tab:nr}), and residuals of the subtraction mask the imprint of $R_3$. In this sample the $(3,2)$
barrier is thus supported by the data in the five runs with $\chi_f\le0.83$ and not constrained in the four with
$\chi_f\ge0.86$; since spin correlates with the leak-to-direct ratio here, this is an outcome of the sample rather than a
spin threshold.

\begin{table*}[t]
\caption{\label{tab:scores}Fixed-prediction scores for the nine runs. $\mathcal{S}_0$, $\mathcal{S}_1$ and
$\mathcal{S}_{\rm real}$: roughness of $\sigma_3$ at $\kappa=0$, $\kappa=1$ and the best real $\kappa$, relative to the
best complex $\kappa$, on the baseline band. Mock: $\kappa=1$ with smooth sources plus measured waveform differences.
Null percentile: fraction (in \%) of the $72$ null pairings ($\tilde h_{32}$ of one run with the $\tilde h_{22}$ of
another) with a smaller value. Ringdown band: $[{\rm Re}\,\omega_{220},\omega_{\rm top}]$. Supported band:
$[0.3,\omega^{\rm sup}_{\rm top}]$, where $\omega^{\rm sup}_{\rm top}$ is the edge of the $(3,2)$ multipole, at which
$|\tilde h_{32}|$ first drops below $30$ times the larger of the floor and the waveform difference ($N=3$ versus $N=2$ for
the first six runs, second versus highest resolution for the third set); the corresponding $(2,2)$ edges are listed
in Table~\ref{tab:manifest}. $^{\dagger}$Best value at the edge of the scan ($|\kappa|=0.5$--$1.5$, $|\arg\kappa|\le0.4$). Last
four rows: roughness of $b_3/R$ at $\kappa=1$ relative to $R=R_3(\chi_f)$.}
\begin{ruledtabular}
\footnotesize
\setlength{\tabcolsep}{1.4pt}
\begin{tabular}{lccccccccc}
 & 0305 & 0540 & 1124 & 4121 & 3128 & 2786 & 0317 & 1443 & 2669\\
\hline
$\mathcal{S}_0$ & 2.24 & 2.01 & 3.31 & 2.11 & 1.49 & 1.44 & 5.23 & 10.04 & 5.09\\
$\mathcal{S}_1$ & 1.02 & 1.27 & 3.29 & 1.73 & 1.23 & 1.08 & 2.44 & 1.50 & 1.52\\
$\mathcal{S}_0/\mathcal{S}_1$ & 2.19 & 1.58 & 1.01 & 1.22 & 1.21 & 1.34 & 2.14 & 6.70 & 3.34\\
best real $\kappa$ & 1.03 & 0.94 & 0.88 & 0.98 & 0.95 & 0.94 & 1.00 & 0.91 & 0.82\\
$\mathcal{S}_{\rm real}$ & 1.00 & 1.16 & 2.53 & 1.70 & 1.16 & 1.03 & 2.44 & 1.24 & 1.16\\
mock: best $|\kappa|$ & 1.01 & 1.01 & 1.00 & 1.00 & 1.00 & 1.01 & 1.00 & 0.99 & 1.04\\
mock: $\mathcal{S}_0$ & 12.3 & 18.7 & 8.7 & 18.6 & 35.7 & 10.7 & 4.4 & 17.3 & 10.6\\
mock: $\mathcal{S}_1$ & 1.05 & 1.06 & 1.00 & 1.00 & 1.00 & 1.11 & 1.06 & 1.00 & 1.07\\
null percentile of $\mathcal{S}_0$ & 97 & 97 & 99 & 97 & 90 & 89 & 99 & 100 & 99\\
null percentile of $\mathcal{S}_0/\mathcal{S}_1$ & 100 & 94 & 69 & 92 & 92 & 93 & 100 & 100 & 100\\
ringdown band: best $\kappa$ & $0.65\,e^{0.40i}$$^{\dagger}$ & $1.08\,e^{0.10i}$ & $0.98\,e^{0.12i}$ & $1.13\,e^{0.05i}$ & $1.50\,e^{0.40i}$$^{\dagger}$ & $0.88\,e^{-0.40i}$$^{\dagger}$ & $0.83\,e^{-0.23i}$ & $0.97\,e^{0.12i}$ & $1.16\,e^{0.40i}$$^{\dagger}$\\
ringdown band: $\mathcal{S}_0$ & 1.11 & 1.38 & 4.33 & 1.49 & 1.31 & 2.47 & 3.43 & 7.25 & 4.11\\
supported band: $M_f\omega^{\rm sup}_{\rm top}$ & 1.27 & 1.12 & 0.98 & 1.03 & 0.95 & 0.95 & 0.51 & 0.55 & 0.61\\
supported band: best $\kappa$ & $1.02\,e^{0.02i}$ & $0.96\,e^{0.07i}$ & $0.97\,e^{0.11i}$ & $0.97\,e^{0.06i}$ & $0.99\,e^{0.05i}$ & $0.97\,e^{0.02i}$ & $0.50\,e^{0.32i}$$^{\dagger}$ & $0.98\,e^{0.07i}$ & $0.97\,e^{0.05i}$\\
supported band: $\mathcal{S}_0$ & 2.54 & 2.21 & 2.81 & 4.01 & 2.88 & 4.20 & 1.07 & 35.77 & 67.04\\
$b_3/R$: no barrier & 3.82 & 2.21 & 0.83 & 0.84 & 0.89 & 0.81 & 1.42 & 4.48 & 3.71\\
$b_3/R$: $R_2$ & 5.41 & 3.46 & 1.25 & 0.85 & 0.85 & 1.15 & 1.73 & 6.59 & 5.55\\
$b_3/R$: $R_3(\chi_f-0.05)$ & 1.06 & 1.37 & 1.19 & 0.83 & 0.84 & 1.06 & 0.93 & 0.74 & 0.56\\
$b_3/R$: $R_3(\chi_f+0.05)$ & 1.45 & 1.00 & 0.81 & 1.37 & 1.24 & 1.04 & 1.15 & 1.73 & 1.65\\
\end{tabular}
\end{ruledtabular}
\end{table*}

\begin{figure*}[t]
\includegraphics[width=\textwidth]{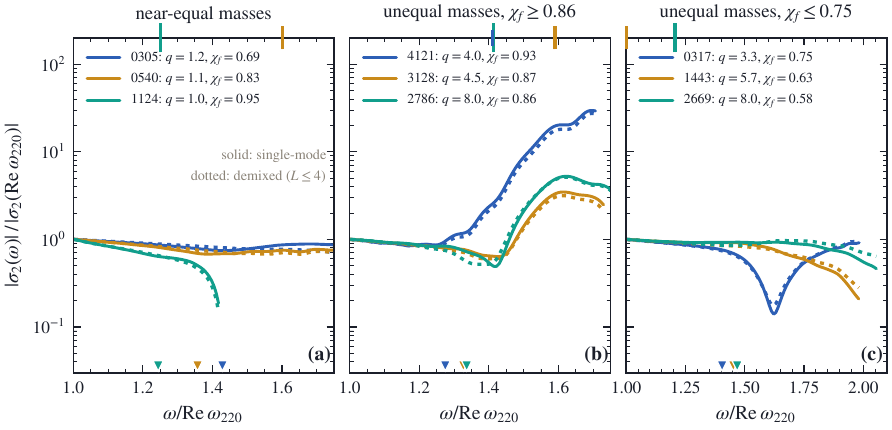}
\caption{\label{fig:src22}Inferred $(2,2)$ source above ${\rm Re}\,\omega_{220}$, normalised there, for (a) the
near-equal-mass mergers and the unequal-mass mergers with (b) $\chi_f\ge0.86$ and (c) $\chi_f\le0.75$. Solid:
single-mode description,
$\tilde h_{22}/(\mu_{22}R_2)$. Dotted: after demixing with spheroidal modes $L=2,3,4$. Each curve ends at the
first frequency where $|\tilde h_{22}|$ drops below $30$ times the floor; ticks on the upper axis mark the upper edge of
the supported band of $\tilde h_{22}$, $\omega^{(22)}_{\rm sup}$ of Table~\ref{tab:manifest} (for SXS:BBH:0305 and
0317 it lies outside the plotted range). Triangles mark ${\rm Re}\,\omega_{320}$ of
each remnant. The rise in (b) is unchanged by demixing; its interpretation is discussed in Sec.~\ref{sec:excess}. The
equal-mass run SXS:BBH:1124 in (a), whose rise grows when its record is shortened, serves as a negative control.}
\end{figure*}

\subsection{Robustness}
\label{sec:robust}
The $\kappa$ test involves five analysis choices: the start of the taper ($-300M_f$ relative to the peak), the length
of the end taper ($40M_f$), the band cut-off ($30$ times the floor), the polynomial degree ($3$), and the lower band
edge ($M_f\omega=0.3$). We varied each in turn: start at $-200M_f$ and $-500M_f$, end taper $20M_f$ and $80M_f$,
cut-off at $10$ and $100$ times the floor, degree $2$, $4$ and $5$, and lower edge $0.45$ and $0.55$, which removes the
frequencies of the late inspiral. For the first four choices and the first six runs we have the baseline plus nine
variants per run, $60$ cases: in $58$ of them the preferred $|\kappa|$ lies between $0.93$ and $1.11$; the exceptions,
$1.15$ and $1.30$, occur for degrees $5$ and $4$ in SXS:BBH:3128 and 2786. The single-mode description $\kappa=0$ is
$1.2$--$4.9$ times rougher than the best $\kappa$ in all $60$ cases. Raising the lower edge leaves the result unchanged
for SXS:BBH:0540, 1124 and 4121 ($|\kappa|=0.96$--$1.01$, $\mathcal{S}_0=1.4$--$3.0$) and for SXS:BBH:0317 and 1443 at
$0.45$, but weakens it elsewhere: for SXS:BBH:0305 the best $|\kappa|$ drops to $0.85$ at $0.45$ and is unconstrained at
$0.55$ ($\mathcal{S}_0=1.07$); for SXS:BBH:3128 it moves to $0.97$ and $1.20$, for SXS:BBH:2786 to $1.14$ and, at
$0.55$, to the edge of the scan, as for SXS:BBH:2669 at $0.55$, with phases up to $\pm0.4$. $\mathcal{S}_0$ stays at
$1.4$--$9.4$ in all runs except SXS:BBH:0305.

\emph{Restricted bands.} On a ringdown-only band, $[{\rm Re}\,\omega_{220},\omega_{\rm top}]$, the single-mode
description remains rougher in eight runs ($\mathcal{S}_0=1.3$--$7.3$), but the preferred $\kappa$ is located near unity
($0.97$--$1.13$) only in SXS:BBH:0540, 1124, 4121 and 1443; in four runs it reaches the edge of the scan
(Table~\ref{tab:scores}). Much of the constraint therefore comes from frequencies below ${\rm Re}\,\omega_{220}$, where
the leak dominates $\tilde h_{32}$. We define the supported band of a multipole as the range in which its amplitude
exceeds $30$ times the larger of the floor and the waveform difference of the NR error budget below ($N=3$ versus $N=2$
for the first six runs, second versus highest resolution for the third set). For the $\kappa$ test, which uses the
band of $\tilde h_{32}$, its upper edge is $M_f\omega=0.95$--$1.27$ for the first six runs; there the preferred $\kappa$ is $0.96$--$1.02$ with phases $0.02$--$0.11$
and $\mathcal{S}_0=2.2$--$4.2$, so the result survives the removal of the least reliable high frequencies. For the third
set only a coarser second resolution exists, whose differences limit this band to $M_f\omega\le0.5$--$0.6$; there
SXS:BBH:1443 and 2669 still give $|\kappa|=0.98$ and $0.97$, while the band of SXS:BBH:0317 is too short to constrain
$\kappa$. At the upper edge of the supported band of $\tilde h_{22}$ ($M_f\omega=0.90$--$1.03$ for the first six runs,
Table~\ref{tab:manifest}) the leak into $\tilde h_{22}$ is $6$--$37\%$,
and the median leak-to-direct ratio in $\tilde h_{32}$ is $0.13$--$8.8$.

\emph{NR error budget.} We repeated the analysis with the $N=3$ and $N=4$ extrapolations of the first six runs and with
the second-highest resolution of seven runs, keeping the remnant parameters fixed. After aligning the peak time and
phase, the extrapolation order changes $|\tilde h_{22}|$ by $0.01\%$--$0.3\%$ at $M_f\omega=0.5$, by $0.3\%$--$9\%$ at
$0.9$ and by $2\%$--$20\%$ at $1.0$; the lower resolution changes it by $0.1\%$--$6\%$ at $0.5$ and by $3\%$ to more than
$100\%$ at $1.0$ (the latter in runs whose second resolution is much coarser). The differences in $\tilde h_{32}$ are
similar. If these differences are added to the floor, the usable band of $\tilde h_{22}$ ends at
$M_f\omega=0.90$--$1.03$ rather than $1.06$--$1.21$ for the first six runs (Table~\ref{tab:manifest}). The QNM ratios change by less than $0.5\%$.
The preferred $|\kappa|$ changes by at most $0.02$, except for SXS:BBH:3128, where all three variants give $0.99$
against $0.94$ at baseline, and for the lower resolution of SXS:BBH:2786 ($1.13$). In summary, at baseline the preferred
$|\kappa|$ lies within $6\%$ of the prediction in seven runs and is $0.91$ and $0.82$ for SXS:BBH:1443 and 2669; over the
variants it ranges from $0.82$ to $1.30$ (Table~\ref{tab:nr}). These ranges reflect analysis and numerical systematics;
a statistical error would require a noise model of the NR data, which we do not attempt.

The preferred phase of $\kappa$ is small but mostly positive: $0.04$--$0.13$ in seven of the nine runs at baseline,
and between $-0.14$ and $0.21$ over all variants. Its sign agrees with that of the phase offset in the QNM-amplitude
ratios (Sec.~\ref{sec:qnmratio}), where the measured phase is $0.03$--$0.13$\,rad less negative than predicted in five
runs. Two analyses of the same waveforms, with the same frame and remnant parameters, thus indicate a small common
phase offset of the $(2,2)$ content of $\tilde h_{32}$ relative to the prediction. Possible origins include the
asymptotic frame of the waveform~\cite{Boyle:2015nqa,Mitman:2022kwt,MaganaZertuche:2021syq}, uncertainties in the
remnant parameters, and contributions not included in Eq.~\eqref{eq:response}, such as second-order or retrograde
modes~\cite{Mitman:2022qdl,Cheung:2022rbm}.

\subsection{The (2,2) multipole at high frequency}
\label{sec:excess}
Figure~\ref{fig:src22} shows the $(2,2)$ source above ${\rm Re}\,\omega_{220}$, inferred with the single-mode
description and after demixing with all spheroidal modes up to $L=4$. For the near-equal-mass runs the source varies by
a factor of $\lesssim2$ over the band; for SXS:BBH:1124 the band ends at $M_f\omega=1.06$, at a deep minimum of
$|\tilde h_{22}|$. For the unequal-mass runs the source reaches a broad minimum at $M_f\omega\simeq0.83$--$0.90$ and then rises
by factors of $5$--$35$, with a corresponding dip and secondary bump in $|\tilde h_{22}|$. Demixing changes this feature
by less than $10\%$, because the leaked $L=3$ reflection is at most $40\%$ of the direct one there. Whatever its origin,
the feature is therefore not spherical--spheroidal mixing as described by Eq.~\eqref{eq:response}. It is also not an
artefact of the extrapolation order or of the resolution at the level these can be tested: with the $N=3$, $N=4$ and
lower-resolution waveforms the rise is $25$--$35$ (SXS:BBH:4121), $5$--$8$ (3128) and $10$--$20$ (2786). The supported
band of the $(2,2)$ multipole (ticks in Fig.~\ref{fig:src22}) contains almost all of the rise of SXS:BBH:3128, a factor
of $5.2$ of $5.5$, and the first factor of $3.0$ of the $35$ of SXS:BBH:4121; the rises of SXS:BBH:2786 and 0317 lie
beyond it. Among the
runs of the third set, SXS:BBH:0317 ($q=3.3$, $\chi_f=0.75$) shows a similar rise, by a factor of $6.5$ above a minimum at
$M_f\omega\simeq0.90$, while for SXS:BBH:1443 and 2669 the usable band ends at $M_f\omega\simeq1.0$, before any rise.

We cannot, however, establish the feature as a physical property of the source, for four reasons. First, it appears where
$|\tilde h_{22}|$ has fallen to $10^{-4}$--$10^{-3}$ of its peak, and dividing by the rapidly falling $|R_2|$
amplifies any residual there. Second, a similar rise, by a factor of about $50$, is present in SXS:BBH:1124 just beyond
its band edge, after the deep minimum of $|\tilde h_{22}|$; when the record of SXS:BBH:1124 is truncated to the
post-peak length of the unequal-mass runs ($144$--$163M_f$, the shortest of the six), this rise grows to factors of
$110$--$230$. Third, varying the end taper between $20$ and $120M_f$ changes the rise to $12$--$45$ (SXS:BBH:4121),
$2.2$--$5.8$ (3128) and $3.8$--$12$ (2786). Fourth, the rise occurs where the waveforms of different extrapolation order and resolution differ by
$5\%$--$50\%$, so its shape, though not its existence, is uncertain. For the three runs of the second set the minimum
also lies close to ${\rm Re}\,\omega_{320}$ of the remnant ($0.90$, $0.86$, $0.85$), where an $\ell=3$ admixture in
$h_{22}$ beyond the spheroidal one, for example from residual centre-of-mass motion or the choice of BMS
frame~\cite{Boyle:2015nqa,Mitman:2022kwt,MaganaZertuche:2021syq}, would interfere with the $(2,2)$ signal; for
SXS:BBH:0317, however, ${\rm Re}\,\omega_{320}=0.78$ lies well below the minimum, and the $(3,2,0)$ amplitude in $h_{22}$
is too poorly determined by time-domain fits to test this directly. Candidate origins therefore include systematic errors of the NR waveforms that the resolution and extrapolation
comparison does not capture~\cite{Knapp:2025ecw} and frame effects. Non-QNM content of the merger signal is a further
possibility, although Ref.~\cite{Cheung:2026gfd} finds that the part of the signal sourced within the prograde light
ring is small, $\lesssim10\%$ of the full signal, and interprets the ``direct wave'' as redistributed QNM power. Smooth-source fits of $|\tilde h_{22}|$ (a power law times $e^{-\omega/T_s}$) leave rms residuals in
$\ln|\tilde h_{22}|$ of $0.07$--$0.12$ for the near-equal-mass runs but $0.24$--$0.44$ for those of the second set,
with or without demixing. The practical conclusion is that greybody-factor fits of $h_{22}$ above $M_f\omega\simeq0.85$
are sensitive to features of the NR data that are not described by any greybody model, and should be checked against
resolution, frame and record length.

\section{Discussion}
\label{sec:discussion}

\begin{figure*}[t]
\includegraphics[width=\textwidth]{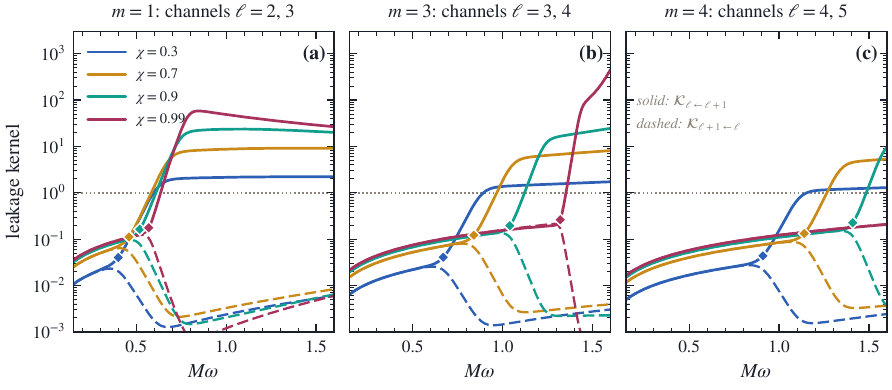}
\caption{\label{fig:otherm}Leakage kernels for (a) $m=1$, (b) $m=3$ and (c) $m=4$, for equal source amplitudes: solid
lines, leak of the reflected $L=\ell+1$ wave into the spherical channel $\ell=\max(2,m)$; dashed lines, leak of $L=\ell$
into the channel $\ell+1$. The dotted line marks equality; diamonds mark ${\rm Re}\,\omega_{\ell m0}$.}
\end{figure*}

\begin{table*}[t]
\caption{\label{tab:manifest}NR run manifest: resolutions (baseline and second), extrapolation orders, remnant mass
$M_f/M$, post-peak record length $T_{\rm post}$, and band edges $\omega_{\rm top}$ of $\tilde h_{22}$ and
$\tilde h_{32}$: $\omega_{\rm top}$, signal $>30$ times the floor; $\omega_{\rm sup}$, signal $>30$ times the larger of
the floor and the waveform difference ($N=3$ versus $N=2$ for the first six runs, second versus highest resolution for
the last three), which defines the supported band (Sec.~\ref{sec:robust}).}
\begin{ruledtabular}
\footnotesize
\begin{tabular}{lcccccccc}
Run & Lev & $N$ & $M_f/M$ & $T_{\rm post}/M_f$ & $M_f\omega_{\rm top}^{(22)}$ & $M_f\omega_{\rm top}^{(32)}$ & $M_f\omega_{\rm sup}^{(22)}$ & $M_f\omega_{\rm sup}^{(32)}$\\
\hline
0305 & 6, 5 & 2, 3, 4 & 0.9520 & 159 & 1.20 & 1.31 & 0.95 & 1.27\\
0540 & 3, 2 & 2, 3, 4 & 0.9323 & 162 & 1.15 & 1.16 & 0.98 & 1.12\\
1124 & 3 & 2, 3, 4 & 0.8828 & 362 & 1.06 & 1.07 & 0.94 & 0.98\\
4121 & 4 & 2, 3, 4 & 0.9461 & 163 & 1.21 & 1.18 & 1.00 & 1.03\\
3128 & 3, 2 & 2, 3, 4 & 0.9604 & 160 & 1.12 & 1.14 & 1.03 & 0.95\\
2786 & 4, 3 & 2, 3, 4 & 0.9764 & 144 & 1.14 & 1.03 & 0.90 & 0.95\\
0317 & 3, 2 & 2 & 0.9642 & 176 & 1.10 & 1.18 & 0.51 & 0.51\\
1443 & 3, 2 & 2 & 0.9802 & 143 & 1.00 & 1.27 & 0.51 & 0.55\\
2669 & 4, 3 & 2 & 0.9861 & 141 & 1.01 & 1.22 & 0.59 & 0.61\\
\end{tabular}
\end{ruledtabular}
\end{table*}

The greybody-factor description of the ringdown rests on the factorisation of the spectrum into a reflection amplitude
and a smooth source. In the spherical basis this factorisation holds only once the mixing is included, either by
demixing the data or by forward-modelling the multipoles with Eq.~\eqref{eq:response}; forward modelling is the safer
choice where a reflection amplitude is deeply suppressed. The spheroidal sources can be recovered by inverting
Eq.~\eqref{eq:response} for pairs of multipoles with the same $m$, using mixing coefficients that are easily
computed~\cite{Stein:2019mop,BHPToolkit}. Our NR tests support this at the level of the angular coefficients: in eight
of nine mergers the range of the fitted $(2,2,0)$ amplitude ratio contains $|\mu_{32}/\mu_{22}|$ (in the ninth it lies
$1$--$4\%$ below). The FD demixing with a fitted strength disfavours the single-mode description beyond the 95th
percentile of a null distribution in seven runs, with the preferred strength within $6\%$ of the prediction at
baseline in seven runs; with the strength fixed at the prediction, the single-mode description is disfavoured at that
level in four runs. The null is not perfectly clean and the mock data are cleaner than the NR spectra, so these
counts are indicative rather than calibrated significances. Both analyses show a small
common phase offset of the $(2,2)$ content of $\tilde h_{32}$, up to $0.13$\,rad at baseline and $0.21$ over the
variants (Table~\ref{tab:nr}), whose origin we have not identified. In this sample the reflection amplitude of the
$(3,2)$ mode is preferred over the alternatives in the five runs with $\chi_f\le0.83$, without resolving the remnant
spin, and is not constrained in the four with $\chi_f\ge0.86$, where the leak dominates; testing it there requires
longer and more accurate NR records or a parametrised source model.

The size of the effect depends on what is being modelled. For the dominant $(2,2)$ multipole of near-equal-mass
mergers, the leak is small near the spectral peak and exceeds the smooth-source fit residuals (rms $0.07$ in
$\ln|\tilde h_{22}|$, Sec.~\ref{sec:excess}) only above $M_f\omega\simeq0.72$, $0.81$ and $0.96$ for the three runs,
where it is partly degenerate with smooth-source parameters; a fit band such as $M\omega=0.35$--$0.75$, used as an
example in Ref.~\cite{Rosato:2025ulx}, is affected only at its upper end. At large mass ratio the $(2,2)$ spectrum above
$M_f\omega\simeq0.85$ carries a feature that mixing does not explain and that we cannot yet attribute
(Sec.~\ref{sec:excess}); greybody-factor fits there should be treated with caution. For the $(3,2)$ multipole the leak
dominates over much of the band, so greybody-factor models of it must include the mixing. We computed the NR comparison
for $m=2$ only. The multipoles fitted in Refs.~\cite{Rosato:2025ulx,Rosato:2026apq}, $(2,1)$, $(3,3)$ and $(4,4)$,
belong to other $m$; their leakage kernels (Appendix~\ref{app:otherm}) are of similar size, and for $(2,1)$ the
$L=3$ leak overtakes the direct reflection already at $M\omega\simeq0.58$--$0.64$. Our results therefore motivate a
check of those templates with the mixing included, but do not measure their bias. The parity prescription of
Sec.~\ref{sec:parity} matters for the $\ell+m$ odd multipoles whenever the relative phases of the sources are fixed by
the model.

Our analysis has several limitations. The sources are assumed smooth in frequency, as in the greybody-factor
literature, and the NR comparison uses nine non-precessing mergers with short post-peak records; the extrapolation-order comparison
uses six of them, and for the other three only a coarser second resolution is available. We compared extrapolation orders and resolutions but not the
dependence on the asymptotic (BMS) frame or on CCE waveforms, which is known to matter at the level of the subdominant
multipoles~\cite{MaganaZertuche:2021syq,Knapp:2025ecw}. In SXS:BBH:1443 and 2669, the runs with the strongest $(3,2)$
sources, the leak into $\tilde h_{22}$ reaches $66\%$ and $38\%$ at the top of the baseline band (although the
resolution comparison supports their spectra only up to $M_f\omega\simeq0.5$--$0.6$), yet the $(2,2)$ channel gives
mixed results; a decisive test of mixing in the dominant multipole needs longer and more accurate records. Precessing
mergers, in which mixing also couples different $m$, are beyond our scope. We worked in linear perturbation theory and
did not include second-order effects~\cite{Mitman:2022qdl,Cheung:2022rbm}, early-time tails, or the $m=0$ and retrograde
content. The deep-tunnelling regime of subdominant modes is limited by the angular projection, which could be improved
with spectral methods in $\theta$.

The scattering method can be extended to modified Teukolsky equations, such as those of the parametrised beyond-Kerr
framework~\cite{Cano:2024jkd}, where FD mode mixing is more involved. This requires, for each equation, the asymptotic
solutions at the horizon and at infinity, a conserved energy flux to normalise the greybody factors, and a rule for
projecting the field onto modes; Ref.~\cite{Cano:2024jkd} provides the equations, not these ingredients. We will
address them in future work.

\FloatBarrier
\section*{Data availability}
The computations used \texttt{pybhpt}~1.1.1~\cite{BHPToolkit},
\texttt{qnm}~0.4.4~\cite{Stein:2019mop} and JAX~0.12.1~\cite{jax2018github}. The NR waveforms are the version-3.0
extrapolated strains of the public SXS catalog~\cite{Boyle:2019kee,Scheel:2025jct}; the runs, resolutions and
extrapolation orders used are listed in Appendix~\ref{app:manifest}. The code and derived data of this work are
available from the authors upon reasonable request.

\section*{Declaration of competing interest}
The authors declare that they have no known competing financial interests
or personal relationships that could have appeared to influence the work
reported in this paper.

\begin{acknowledgments}
T. Tangphati acknowledges COST action CA21106 and CA22113. We thank the SXS Collaboration for making their simulations public, and the developers of the Black Hole Perturbation
Toolkit and of the \texttt{qnm} package.
\end{acknowledgments}

\appendix

\section{Ingoing multipole data}
\label{app:data}
In flat space the $s=-2$ Teukolsky equation for $\Psi=\rho^{-4}\psi_4$ with angular dependence $\Yw_{\ell m}$ admits
purely ingoing solutions that are finite sums $\Psi=\sum_{j=0}^{\ell-2}c_j F^{(\ell-2-j)}(v)/r^{j+1}$, with $v=t+r$
and $F$ arbitrary. Substituting this ansatz and requiring every power of $1/r$ to vanish gives, with $c_0=1$, the
recurrence
\begin{equation}
\frac{c_{j+1}}{c_j}=\frac{(j+1)(j+4)-\lambda_\ell}{2(j+1)},\qquad \lambda_\ell=\ell(\ell+1)-2,
\end{equation}
which terminates at $j=\ell-2$ and yields Eq.~\eqref{eq:indata} for $\ell=2,3,4$. On the initial slice $\tau=0$ we set
$\psi=r\Delta^{-2}\Psi$ and $\partial_\tau\psi=r\Delta^{-2}\partial_\tau\Psi$, where $\partial_\tau\Psi$ follows from
$\Psi$ by replacing each $F^{(n)}(v)$ with $F^{(n+1)}(v)$, with $v$ the ingoing null coordinate evaluated on the slice.
The hyperboloidal variable $\psi$ is suppressed by $r^{-4}$ for ingoing waves at large $r$, which is harmless in a linear
problem. We set $\psi$ and $\partial_\tau\psi$ to zero inside the horizon and at $\scri$ initially.

\section{Numerical scheme}
\label{app:numerics}
Near the horizon and at $\scri$ the coefficients of the hyperboloidal equation involve cancelling factors of $1/\Delta$
and of $\sqrt{1+r^2}-r$; we evaluate them there in 80-digit arithmetic and elsewhere in double precision. Radial and
polar derivatives are fourth-order finite differences, one-sided at the two outflow boundaries. The second radial
derivative is computed as $D_1(D_1\psi)$, because the compact second-derivative stencil lets grid-scale modes grow near
the horizon, where the principal coefficient changes sign. The polar grid is cell-centred, with parity $(-1)^{m+s}$
across the poles. Time integration uses the classical fourth-order Runge--Kutta method with Courant factor $0.5$
relative to the largest characteristic speed. We add sixth-derivative Kreiss--Oliger dissipation in both directions,
with strength $\epsilon_{\rm KO}=0.05$ at the horizon falling to $10^{-3}$ away from it (tanh profile of width $0.08$ in
$X$); the strengths scale with the grid spacing, so the dissipation vanishes in the continuum limit. The scheme
converges at fourth order: for $\chi=0$ and $m=2$, over the first $50M$ of evolution the error ratio for $201/401/801$
radial points is $16.0$ both at the horizon and at $\scri$. Late-time tails of compactly supported data reach the
double-precision round-off floor in our runs; they play no role in this paper.

\section{Normalisation of the horizon flux}
\label{app:norm}
At $\chi=0$ we fitted the time-domain greybody factor~\eqref{eq:Gtd} with the horizon flux multiplied by a free
constant $n_H$ (in place of $1/2\pi$) against the Regge--Wheeler result over $0.25\le M\omega\le0.8$. The fit gives
$2\pi n_H=0.99998$, i.e.\ $n_H=0.159152$ against $1/(2\pi)=0.159155$. This confirms the normalisation and the frequency dependence of
the horizon flux formula of Ref.~\cite{Sasaki:2003xr}. All results in the paper use the published formula without a
fitted constant.

\section{Regge--Wheeler and Zerilli reflection amplitudes}
\label{app:rwz}
We integrate $\dd^2\mathcal{X}/\dd r_*^2+(\omega^2-V)\mathcal{X}=0$ for the Regge--Wheeler and Zerilli
potentials~\cite{Regge:1957td,Zerilli:1970se} from $r=2M(1+5\times10^{-6})$ with $\mathcal{X}=e^{-i\omega r_*}$, using
$r_*=r+2M\ln(r/2M-1)$ as in the Teukolsky solutions, to $r=400M$ with relative tolerance $10^{-12}$. There we match to
$\mathcal{X}=A_{\rm out}u_++A_{\rm in}u_-$, where $u_\pm=e^{\pm i\omega r_*}\sum_{n=0}^{10}a^{\pm}_n r^{-n}$ are asymptotic
series whose coefficients follow from the recursion obtained by substituting them into the equation. The reflection
amplitude is $R=A_{\rm out}/A_{\rm in}$ and the transmission is $|1/A_{\rm in}|^2$. The horizon condition is imposed at
leading order only, without higher terms of the near-horizon expansion; this, rather than the integration tolerance
($10^{-12}$), limits the accuracy of $R_{\rm RW}$ and $R_{\rm Z}$ to about $10^{-4}$. The agreement
quoted after Eq.~\eqref{eq:parity} is the maximum over the tested frequencies of the complex relative difference
$|R_L/R_{\rm Z,RW}-1|$; for $\ell=2$ its value, $4.1\times10^{-4}$, is at this level.

\section{Leakage kernels for $m=1$, $3$ and $4$}
\label{app:otherm}
Figure~\ref{fig:otherm} shows the leakage kernels $\mathcal{K}_{\ell\leftarrow\ell+1}$ and
$\mathcal{K}_{\ell+1\leftarrow\ell}$ for the pairs $(2,1)$--$(3,1)$, $(3,3)$--$(4,3)$ and $(4,4)$--$(5,4)$, computed as
in Sec.~\ref{sec:mixing}. For $m=1$ the $L=3$ leak overtakes the direct $L=2$ reflection already at
$M\omega=0.60$, $0.58$, $0.61$ and $0.64$ for $\chi=0.3$, $0.7$, $0.9$ and $0.99$, and exceeds it by factors of $2.2$,
$8.7$, $24$ and $49$ at $M\omega=1$. For $m=3$ the crossing lies at $M\omega=0.89$, $0.97$, $1.12$ and $1.35$, close to
${\rm Re}\,\omega_{330}$ at high spin, and for $m=4$ at $M\omega=1.15$, $1.27$ and $1.49$ for $\chi=0.3$, $0.7$ and $0.9$
(above $M\omega=1.6$ for $\chi=0.99$). Well below the respective fundamental QNM frequency
($M\omega<0.6\,{\rm Re}\,\omega_{\ell m0}$) the reverse leak into the upper channel is $0.5$--$13\%$. As for $m=2$, the effect on an NR multipole is the kernel times the source ratio.

\section{NR run manifest}
\label{app:manifest}
Table~\ref{tab:manifest} lists the SXS waveforms used. All are catalog version 3.0, extrapolated strain; the baseline
uses extrapolation order $N=2$ at the highest resolution. Frequencies are in units of the remnant mass $M_f$ from the
catalog metadata.

\bibliographystyle{apsrev4-2}
\bibliography{refs}

\end{document}